\documentclass[twocolumn,showpacs,aps,floatfix,prd,nofootinbib]{revtex4}

\usepackage{graphicx}
\usepackage{dcolumn}
\usepackage{amsmath}
\usepackage{epsfig}
\usepackage{colordvi}
\usepackage{color}
\usepackage{hhline}

\RequirePackage{xspace}

\newcommand{\BABARPubYear}    {06}
\newcommand{\BABARPubNumber}  {024}
\newcommand{\SLACPubNumber} {11849}

\usepackage{relsize}
\def\babar{\mbox{\slshape B\kern-0.1em{\smaller A}\kern-0.1em
    B\kern-0.1em{\smaller A\kern-0.2em R}}}
\def\gev   {\ensuremath{\mathrm{\,Ge\kern -0.1em V}}\xspace}
\def\mev    {\ensuremath{\mathrm{\,Me\kern -0.1em V}}\xspace}
\def\mevc  {\ensuremath{{\mathrm{\,Me\kern -0.1em V\!/}c}}\xspace}
\def\gevc  {\ensuremath{{\mathrm{\,Ge\kern -0.1em V\!/}c}}\xspace}
\def\mevcc {\ensuremath{{\mathrm{\,Me\kern -0.1em V\!/}c^2}}\xspace}
\def\gevcc {\ensuremath{{\mathrm{\,Ge\kern -0.1em V\!/}c^2}}\xspace}
\mathchardef\Upsilon="7107
\def\Y#1S{\ensuremath{\Upsilon{(#1S)}}\xspace}

\def\pep2{PEP-II}
\def\gisr     {\ensuremath {\gamma_{\scriptscriptstyle ISR}} \xspace}
\def\etappp   {\ensuremath {\eta^{(\prime)}} \xspace}
\def\eetoeg   {\ensuremath {e^+e^- \!\!\to\! \eta\gamma} \xspace}
\def\eetoepg  {\ensuremath {e^+e^- \!\!\to\! \eta^\prime\gamma} \xspace}
\def\eetoeppg {\ensuremath {e^+e^- \!\!\to\! \etappp\gamma} \xspace}
\def\epem     {\ensuremath{e^+e^-}\xspace}     
\def\chipppng {\ensuremath{\chi^2_{2\pi\pi^0\gamma}}\xspace}
\def\chippeg  {\ensuremath{\chi^2_{2\pi\eta\gamma}}\xspace}
\def\chifppng {\ensuremath{\chi^2_{4\pi\pi^0\gamma}}\xspace}
\def\BR         {{\ensuremath{\cal B}\xspace}}
\def\pip   {\ensuremath{\pi^+}\xspace}
\def\pim   {\ensuremath{\pi^-}\xspace}

\long\def\inst#1{\par\nobreak\kern 4pt\nobreak
  {\it #1}\par\vskip 10pt plus 3pt minus 3pt}
  
\begin{document}

\begin{flushleft}
SLAC-PUB-\SLACPubNumber \\
\babar-PUB-\BABARPubYear/\BABARPubNumber \\
\end{flushleft}

% Title of the paper
\title{\large \bf \boldmath
Measurement of the $\eta$ and $\eta^\prime$ transition form factors at
$q^2=$112$\gev^2$
} % end title
\author{B.~Aubert}
\author{R.~Barate}
\author{M.~Bona}
\author{D.~Boutigny}
\author{F.~Couderc}
\author{Y.~Karyotakis}
\author{J.~P.~Lees}
\author{V.~Poireau}
\author{V.~Tisserand}
\author{A.~Zghiche}
\affiliation{Laboratoire de Physique des Particules, 
F-74941 Annecy-le-Vieux, France }
% ...
\collaboration{The \babar\ Collaboration}
\noaffiliation

\date{\today}

% Abstract
\begin{abstract}
We report a study of the processes \eetoeg and \eetoepg at a
center-of-mass energy of 10.58~\gev,
using a 232~fb$^{-1}$ data sample collected with the \babar\ detector
at the \pep2 collider at SLAC.
We observe 
$20^{+6}_{-5}$ $\eta\gamma$ and 
$50^{+8}_{-7}$ $\eta^\prime\gamma$ events over small backgrounds, and
measure the cross sections
$\sigma(\eetoeg) =4.5^{+1.2}_{-1.1}\pm 0.3$~fb and
$\sigma(\eetoepg)=5.4  \pm     0.8 \pm 0.3$~fb.
The corresponding transition form factors at $q^2\! =\! 112~\gev^2$ are
$q^2|F_\eta(q^2)|=0.229\pm0.030\pm0.008$~\gev, and
$q^2|F_{\eta^\prime}(q^2)|=0.251\pm0.019\pm0.008$~\gev, respectively.
\end{abstract}


\maketitle

\setcounter{footnote}{0}

\section{ \boldmath Introduction}\label{intro}

The cross section for the reaction 
$e^+e^- \!\!\to\! \gamma^\ast \!\!\to\! P\gamma$,
where $P$ is a pseudoscalar meson, is given, for energies large
compared with the $P$ mass $m_P$, by
\begin{equation}
\frac{d\sigma_{e^+e^-\to P\gamma}(s,\theta^\ast_\gamma)}
     {d\cos{\theta^\ast_\gamma}} =
\frac{\pi^2\alpha^3}{4}|F_P(s)|^2 (1+\cos^2 \theta^\ast_\gamma),
\label{eq1}
\end{equation}
where $\sqrt{s}$ is the $e^+e^-$ center-of-mass (c.m.) energy, 
$\theta^\ast_\gamma$ is the angle between the outgoing photon and the
incoming electron in the $e^+e^-$ c.m.\ frame, 
and $\alpha$ is the fine structure constant.
The form factor $F_P(q^2)$ describes the effect of the strong
interaction on the $\gamma^\ast \!\to \gamma P$ transition as a
function of the four-momentum $q$ of the virtual photon;
here $q^2\! =\! s$.

These transition form factors can be calculated using perturbative
Quantum Chromodynamics (QCD) in the asymptotic limit, 
$q^2\! >\! > \! m^2_P$~\cite{Brodsky,Braaten}:
\begin{equation}
-q^2F_P(q^2)=\sqrt{2}f_P\left(1-\frac{5}{3}\frac{\alpha_s(q^2)}{\pi}\right),
\label{eq3}
\end{equation}
where $f_P$ is the pseudoscalar meson decay constant, and $\alpha_s$
is the strong coupling.
The $\pi$-meson decay constant is known from leptonic $\pi$ decays to be
about 131~\mev. 
The effective $\eta$ and $\eta^\prime$ decay constants depend on the
mixing between the two states, which must be calculated from other
data~\cite{Feldman,Kaiser,Benayoun,Goity,DeFazio,Escribano};
for example, the scheme in Ref.~\onlinecite{Feldman} gives
$f_\eta \!\approx\! f_\pi$ and 
$f_{\eta^\prime} \!\approx\! 1.6f_\pi$~\cite{Kroll}.
At lower $q^2$, however, the form factor can only be estimated
phenomenologically.
Currently, measurements of \eetoeg cover only 
the energy region below $\sqrt{s}\!=\!1.4$~\gev~\cite{SND,CMD2}, 
where decays of $\rho(770)$, $\omega(782)$, and $\phi(1020)$ dominate.
There are also measurements from $J/\psi \!\!\to\! \eta\gamma$ and from
$\phi, J/\psi, \psi(2S) \!\!\to\! \eta^\prime\gamma$ decays~\cite{pdg}.
Space-like $\etappp\gamma$ transition form factors have been measured in
two-photon reactions 
$\gamma\gamma^\ast \!\to \etappp$~\cite{CLEO97,L397,CELLO91,TPC90,PLUTO}
up to $q^2 \!\approx\! 20~\gev^2$.
These $q^2$ values are not in the asymptotic region,
and measurements at higher $q^2$ are needed both to establish the
asymptotic value and to test phenomenological models.

In this article we present measurements of the reaction \eetoeppg at an
average $e^+e^-$ c.m. energy of 10.58~\gev, 
corresponding to $q^2\! =\! 112~\gev^2$.
We reconstruct the $\eta$ in the $\pi^+\pi^-\pi^0$ decay mode, and the
$\eta^\prime$ in the $\pi^+\pi^-\eta$ decay mode, where the
intermediate $\eta$ state decays to either $\gamma\gamma$ or
$\pi^+\pi^-\pi^0$.
From Eqs.(\ref{eq1}) and~(\ref{eq3}) and the $f_P$ values given above, 
we expect cross sections of
\begin{eqnarray}
\sigma(e^+e^- \!\to \eta\gamma)        & \approx & 2.1\mbox{ fb}
\label{xsexp} \\
\sigma(e^+e^- \!\to \eta^\prime\gamma) & \approx & 5.5\mbox{ fb}, \nonumber
\end{eqnarray}
which are much smaller than those of many hadronic processes, 
so we must consider other sources of such events, as well as
backgrounds, carefully.
About 20\% of the hadronic events in our data are from decays of
the \Y4S resonance;
its branching fraction into $\etappp\gamma$ has not been measured,
but can be estimated using the relation 
\begin{equation}
\frac{\Gamma(\Y4S \to \etappp\gamma)}{\Gamma(\Y1S \to \etappp\gamma)}
\approx
\frac{\Gamma(\Y4S \to e^+e^-)}       {\Gamma(\Y1S \to e^+e^-)}.
\label{gratio}
\end{equation}
From the upper limit on the branching fraction
$\BR(\Y1S \!\!\to\! \etappp\gamma)\! <\,$2.1(1.6)$\times 10^{-5}$ at
90\% CL~\cite{pdg}, we obtain 
$\BR(\Y4S \!\!\to\! \etappp\gamma)\! <\,$2.5(1.9)$\times 10^{-8}$ and
a cross section,
$\sigma(e^+e^- \!\!\to\! \Y4S \!\!\to\! \etappp\gamma)\! <\,$0.026(0.020)~fb,
well below the values expected for the mechanism under study.
Radiative return, 
$e^+e^- \!\!\to\! \gisr e^+e^- \!\!\to\! \gisr\etappp$,
in which there is a high energy photon \gisr from initial-state
radiation (ISR) off the initial electron or positron, 
is forbidden in single-photon annihilation of the resulting $e^+e^-$ pair.
Double-photon exchange is estimated to have a cross section much smaller than
in Eqs.(\ref{xsexp})~\cite{Chernyak}.                         
We therefore assume all the true \eetoeppg events in the
data are due to the processes under study.

The radiative processes 
$e^+e^- \!\!\to\! \gisr \pi^+\pi^-\pi^0$ and
$e^+e^- \!\!\to\! \gisr \pi^+\pi^-\eta$ produce final states identical to
those for the signals.
However, the $\pi^+\pi^-\pi^0$ and $\pi^+\pi^-\eta$ mass distributions
for these processes do not show peaks at the $\eta$ or $\eta^\prime$
masses,
and we include this background in the fits to the mass distributions.
Other sources of non-peaking background, such as higher multiplicity
ISR events with missing particles and $e^+e^- \!\!\to\,$hadrons events
with a high energy $\pi^0$ faking a hard photon, are reduced to low
levels in the selection process.

Background that peaks in the \etappp mass region
arises mainly from the ISR processes
$e^+e^- \!\!\to\! \gisr V \!\!\to\! \gisr\etappp\gamma$, 
where $V$ is a vector meson, 
such as $\rho$, $\omega$, $\phi$, $J/\psi$ or $\Upsilon$.
If the photon from the vector meson decay has low energy in the
laboratory frame and is lost, the event cannot be distinguished from
signal. 
Additional peaking background can arise from
$e^+e^- \!\!\to\! VP \!\!\to\! \etappp\pi^0\gamma$, with or without an
ISR photon,
where $V$ is a vector meson decaying into $\pi^0\gamma$,
$\eta\gamma$ or $\eta^\prime\gamma$, and $P$ is a $\pi^0$, $\eta$, or
$\eta^\prime$. 
These backgrounds are estimated from Monte Carlo (MC) simulation and data,
and are subtracted from the number of observed $\etappp\gamma$ events.

\section{ \boldmath The \babar\ detector and data samples}
\label{detector}

Here we analyze a data sample of 232 fb$^{-1}$ collected with the 
\babar\ detector~\cite{ref:babar-nim} at the \pep2\ facility, where
9.0~\gev electrons collide with 3.1~\gev positrons at a c.m.\ energy of 
10.54--10.58~\gev.
Charged-particle tracking is
provided by the five-layer silicon vertex tracker (SVT) and
the 40-layer drift chamber (DCH), operating in a 1.5~T axial
magnetic field. 
The transverse momentum resolution is 0.47\% at 1~\gevc. 
Energies of photons and electrons are measured with a CsI(Tl) 
electromagnetic calorimeter (EMC) with a resolution of 3\% at 1~\gev. 
Charged-particle identification is provided by ionization measurements in
the SVT and DCH, and by an internally reflecting ring-imaging
Cherenkov detector (DIRC).
Full detector coverage is available over the polar angle range
30$^\circ<\theta^\ast<$140$^\circ$ in the c.m.\ frame.

We simulate the signal processes using a MC
generator based on Eq.(\ref{eq1}). 
The simulation of ISR background processes uses two methods:
the Bonneau-Martin formula~\cite{BM} for 
$\epem \!\!\to\! \gisr V$, with 
$V\! =\! \rho,\omega,\phi,J/\psi \!\to\! \etappp\gamma$,
and $\epem \!\!\to\! \gisr\omega\etappp$, with
$\omega \!\!\to\! \pi^0\gamma$;
and the more accurate approach developed in Ref.~\onlinecite{ckhhad}
where the hadron angular distributions are important, for
$e^+e^- \!\!\to\! \gisr 3\pi$, 
$e^+e^- \!\!\to\! \gisr\pi^+\pi^-\eta$ and 
$e^+e^- \!\!\to\! \gisr 4\pi$.
Since the polar angle distribution of the ISR photon peaks near
$0^\circ$ and $180^\circ$, we generate events only over the range
20$^\circ\! < \theta^\ast_\gamma\! <$160$^\circ$,
except for $e^+e^- \!\!\to\! \gisr\Upsilon(nS)$, where
the ISR photon is generated over the full polar angle range.
We also simulate non-ISR events of the type 
$e^+e^- \!\!\to\! \omega\etappp$ with $\omega \!\!\to\! \pi^0\gamma$.
We simulate extra soft-photon radiation from the initial state in all cases
using the structure function method of Ref.~\onlinecite{strfun}, with
the extra photon energy restricted such that the invariant mass 
of the hadronic (plus ISR photon) system must exceed 8~\gevcc for
non-ISR (ISR) processes.
We study backgrounds from $e^+e^- \!\!\to\! q\bar{q}$ using the 
JETSET~\cite{Jetset} package. 

We simulate the detector response, including interactions of the 
generated particles with the detector material, using the 
GEANT4~\cite{ref:geant4} package, taking
into account the variation of the detector operating conditions with time.
We simulate the beam-induced background, 
which may lead to the appearance of extra photons and tracks in 
the events of interest, by overlaying the raw data from a random
trigger event on each generated event.

\section{ \boldmath Event selection}
\label{selection}

The initial selection of events requires the presence of
a high-energy photon with momentum roughly opposite to the vector sum of the
good-quality charged tracks and other photons.
The hard photon must have energy in the c.m.\ frame $E^\ast_\gamma>3$~\gev;
charged tracks must extrapolate to the interaction region,
have a momentum transverse to the beam direction above 100~\mevc, 
and have a polar angle in the laboratory frame in the region 
 23$^\circ\! <\! \theta\! <\! 140^\circ$
(38$^\circ\! <\! \theta^\ast\! <\! 154^\circ$ in the c.m.\ frame).

We study the \eetoeg and \eetoepg reactions in the $\pi^+\pi^-3\gamma$ 
and $\pi^+\pi^-\pi^+\pi^-3\gamma$ final states,
i.e. we use the $\eta \!\!\to\! \pi^+\pi^-\pi^0$ decay mode for the former and
the $\eta^\prime \!\!\to\! \eta\pi^+\pi^-$ mode, 
with $\eta \to \gamma\gamma$ and $\eta \!\!\to\! \pi^+\pi^-\pi^0$, for
the latter.
Since a significant fraction of the events contain beam-generated 
spurious tracks and photon candidates, 
we select events with at least two (four) tracks and at least three 
photons with energies above 100~\mev (50~\mev) for the $2\pi 3\gamma$ 
($4\pi 3\gamma$) final state.

\begin{figure*}
\includegraphics[width=0.325\linewidth]{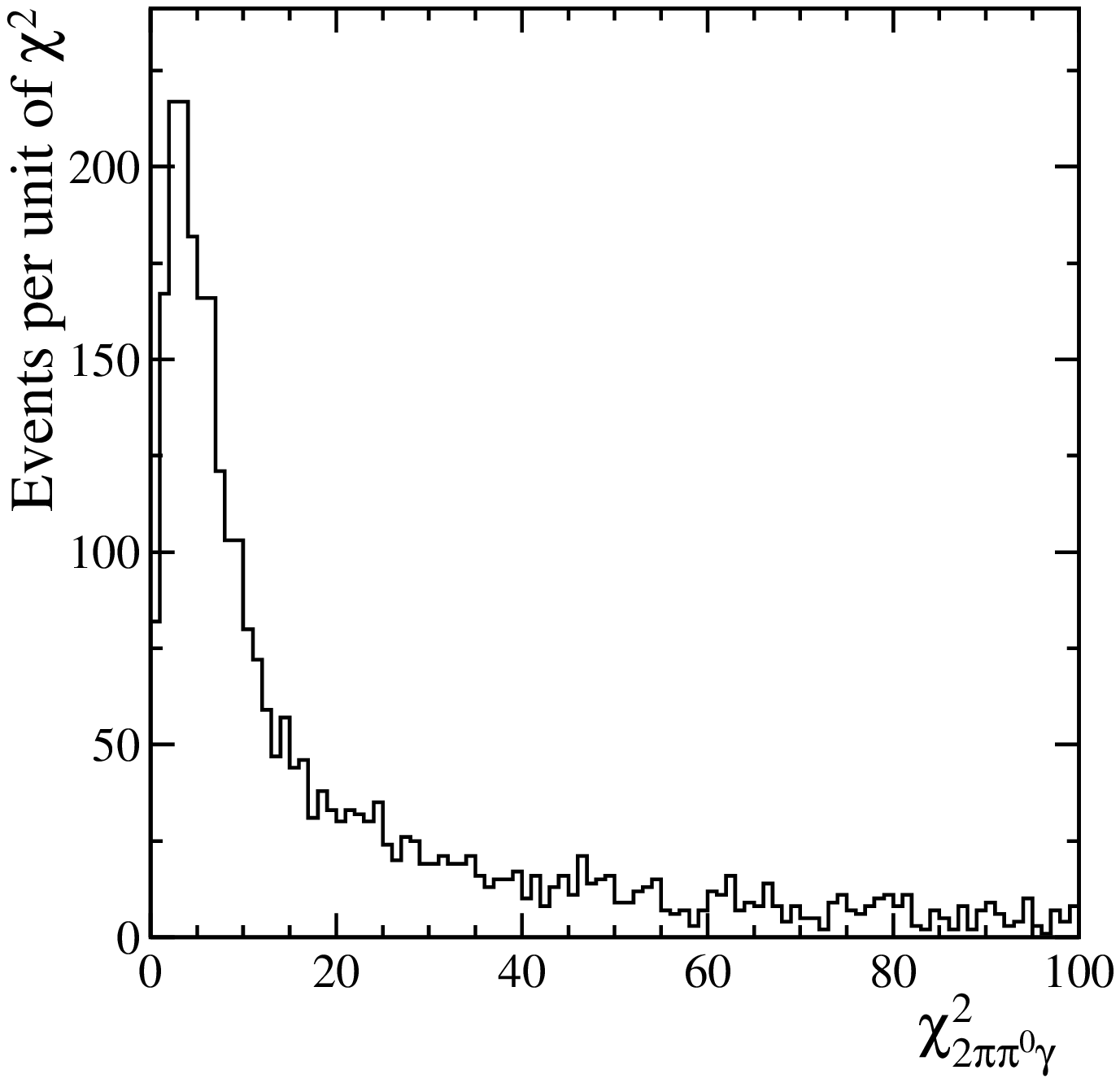}
\includegraphics[width=0.325\linewidth]{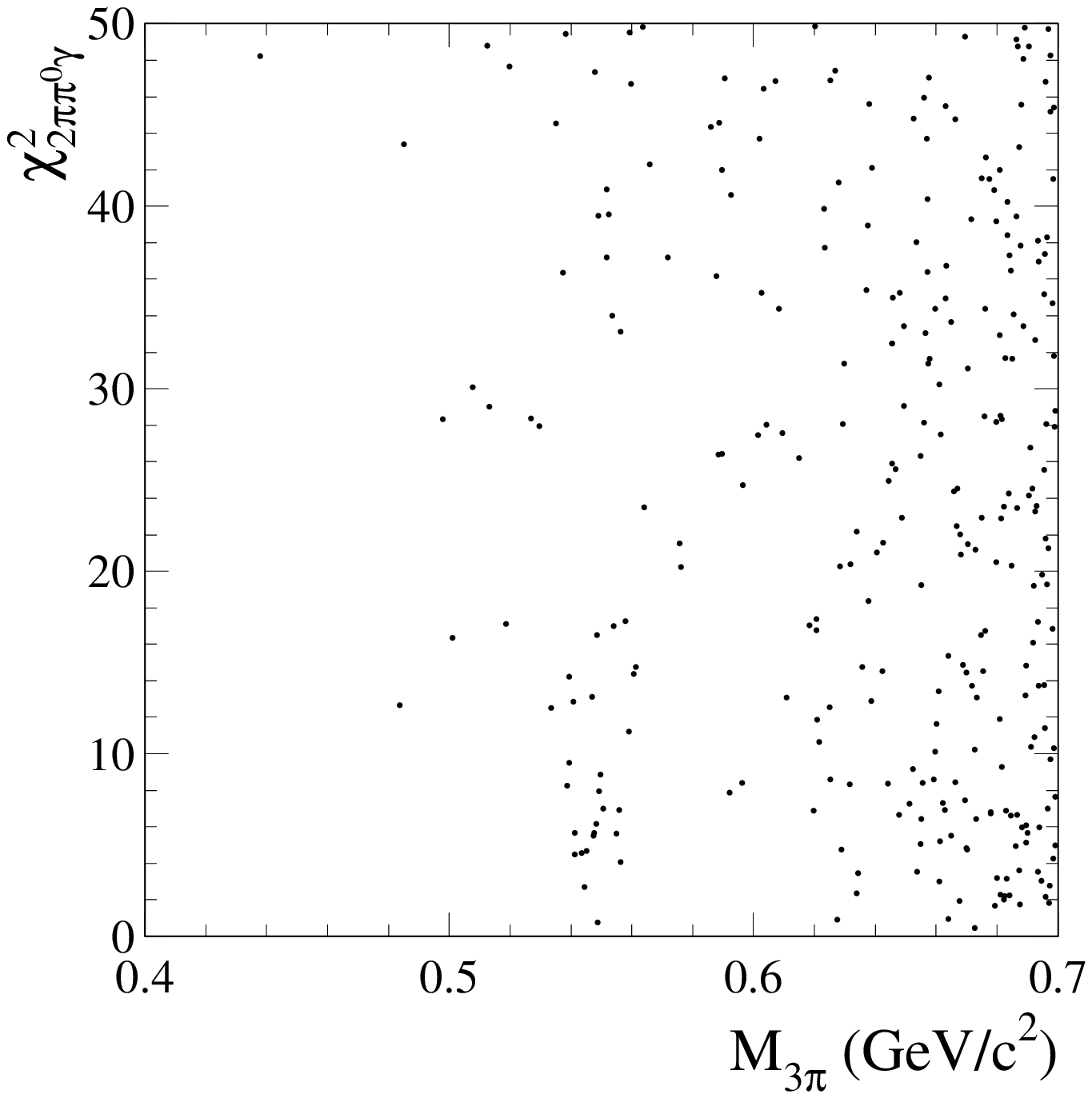}
\includegraphics[width=0.325\linewidth]{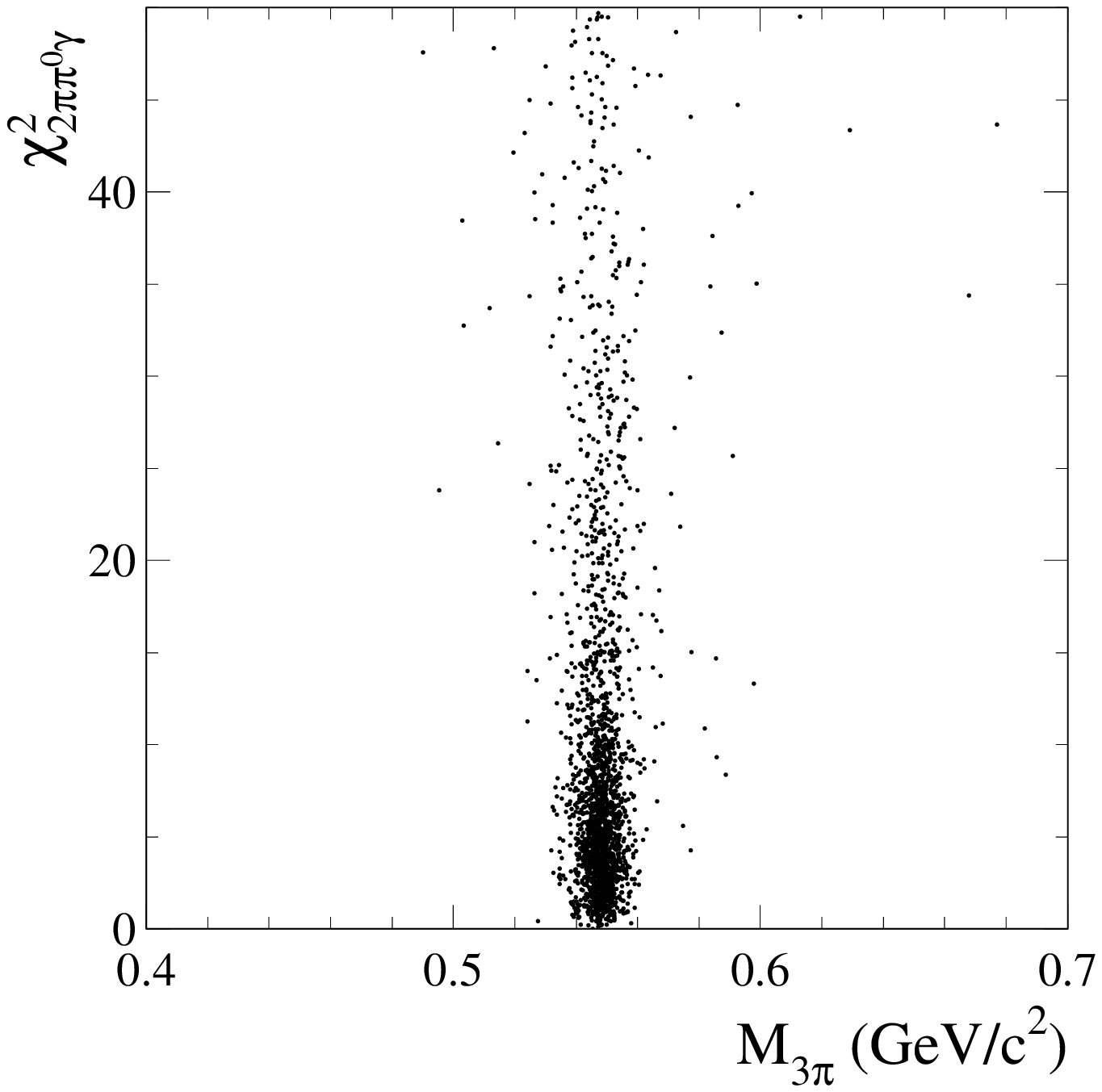}
\vspace{-0.2cm}
\caption{
  Distribution (left) of \chipppng for simulated 
  $\eetoeg \!\!\to\! \pip\pim\pi^0\gamma$ signal events.
  Scatter plots of \chipppng versus the $\pip\pim\pi^0$ invariant mass for the
  selected events in data (center) and signal simulation (right).
  }
\label{chim}
\end{figure*}

We assume the photon with the highest 
$E^{\ast}_\gamma$ is the recoil photon, and
consider only the set of two or four tracks with zero total charge
that has the smallest sum of distances from the interaction point in the
azimuthal plane.
We fit a vertex to this set of tracks, 
which is used as the point of origin to calculate all photon angles.
We accept pairs of other photons as $\pi^0$ or $\eta$ candidates
if their invariant mass is in the range 0.07--0.20~\gevcc or
0.45--0.65~\gevcc, respectively.
For each such candidate, we perform a kinematic fit to 
the selected tracks and photons that
imposes energy and momentum conservation and constrains the candidate
$\pi^0$ or $\eta$ invariant mass.
We use the $\chi^2$ of the kinematic fit
(\chipppng, \chippeg or \chifppng)
to discriminate signal from background.
The simulation does not reproduce the shape of the photon energy resolution
function, especially at high energy.
Since this distorts the $\chi^2$ distributions,
only the measured direction of the recoil photon is
used in the fit; its energy is a free parameter.
For events with more than one $\pi^0$ and/or $\eta$ candidate,
the one giving the lowest $\chi^2$ is retained.
The distribution of $\chi^2_{2\pi\pi^0\gamma}$ for simulated
\eetoeg events is shown in Fig.~\ref{chim}a.
There are four effective degrees of freedom, and the distribution shows a
long tail due to higher-order photon radiation.
We also perform the kinematic fit without the mass constraint,
calculate $\chi^2_{2\pi 3\gamma}$,
and use the $\chi^2$ difference 
($\chipppng \! -\!\chi^2_{2\pi 3\gamma}$ or
 $\chippeg  \! -\!\chi^2_{2\pi 3\gamma}$)
as a measure of the $\pi^0$ or $\eta$ reconstruction quality.

\begin{figure}[hbt]
\begin{center}
\includegraphics[width=0.8\linewidth]{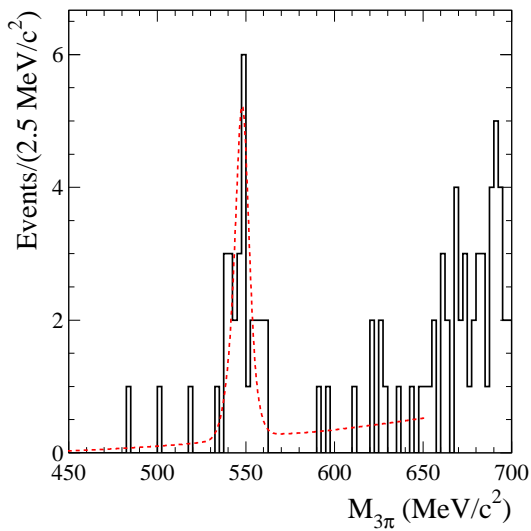}
\vspace{-0.2cm}
\caption{
  The $3\pi$ invariant mass spectrum for the \eetoeg candidates in the
  data.
  The dashed curve represents the result of the fit described in the text.
  }
\label{etafit}
\end{center}
\end{figure}

To suppress backgrounds in the \eetoeg sample from events containing
kaons and events from multi-particle ISR, QED, and $e^+e^- \!\to\! q\bar{q}$ 
processes, while maintaining high signal efficiency,
we consider events with exactly one pair of selected tracks and 
no more than one additional track.
Considering the selected pair, we require that:
i) neither track is identified as a kaon;
ii) $(E_1 / p_1)\! +\! (E_2 / p_2)\! <\! 1.5$,
where $E_i$ is the EMC energy deposition associated with the $i^{th}$ track and
$p_i$ is its measured momentum;
iii) $\chipppng \!\! -\! \chi^2_{2\pi 3\gamma}\! <\,$5;
and 
iv) the invariant mass of the two charged tracks $M_{2\pi}\! <\! 415~\mevcc$.
Requirement (ii) suppresses dielectron events;
requirement (iv) only suppresses background events with a 
$\pi^+\pi^-\pi^0$ mass $M_{3\pi}\! >\,$0.6~\gevcc, but it facilitates the
extrapolation of the background under the $\eta$ peak.

We show scatter plots of \chipppng versus $M_{3\pi}$
for the selected candidates in the data and the \eetoeg signal
simulation in Figs.~\ref{chim}b and~\ref{chim}c.
A cluster of data events is evident near the $\eta$ mass at
small values of \chipppng.
Figure~\ref{etafit} shows the $M_{3\pi}$ distribution for data events
with $\chipppng\! <\! 20$.
In order to determine the number of events containing a true $\eta$ we
perform a binned maximum likelihood fit to the
$M_{3\pi}$ spectrum over the range 450--650~\mevcc with a sum of signal
and background distributions.
We describe the signal by a sum of three Gaussian functions with
parameters obtained from the simulation, convolved with an additional
Gaussian smearing function of width $\sigma_G\! =\! 1.3^{+0.6}_{-1.0}$~\mevcc
determined from high-statistics $\omega \!\!\to\! \pip\pim\pi^0$ data 
(see Sec.~\ref{deteff}).
The background is a second order polynomial.
The line on Fig.~\ref{etafit} represents the result of the fit.
The fitted number of events is $N_\eta\! =\! 22.7^{+5.6}_{-4.9}\pm0.6$,
where the first error is statistical and the second is the systematic
arising from the uncertainty on $\sigma_G$, variation of the
background parameters, and using a first or third order
polynomial background.

\begin{figure}
\includegraphics[width=0.7\linewidth]{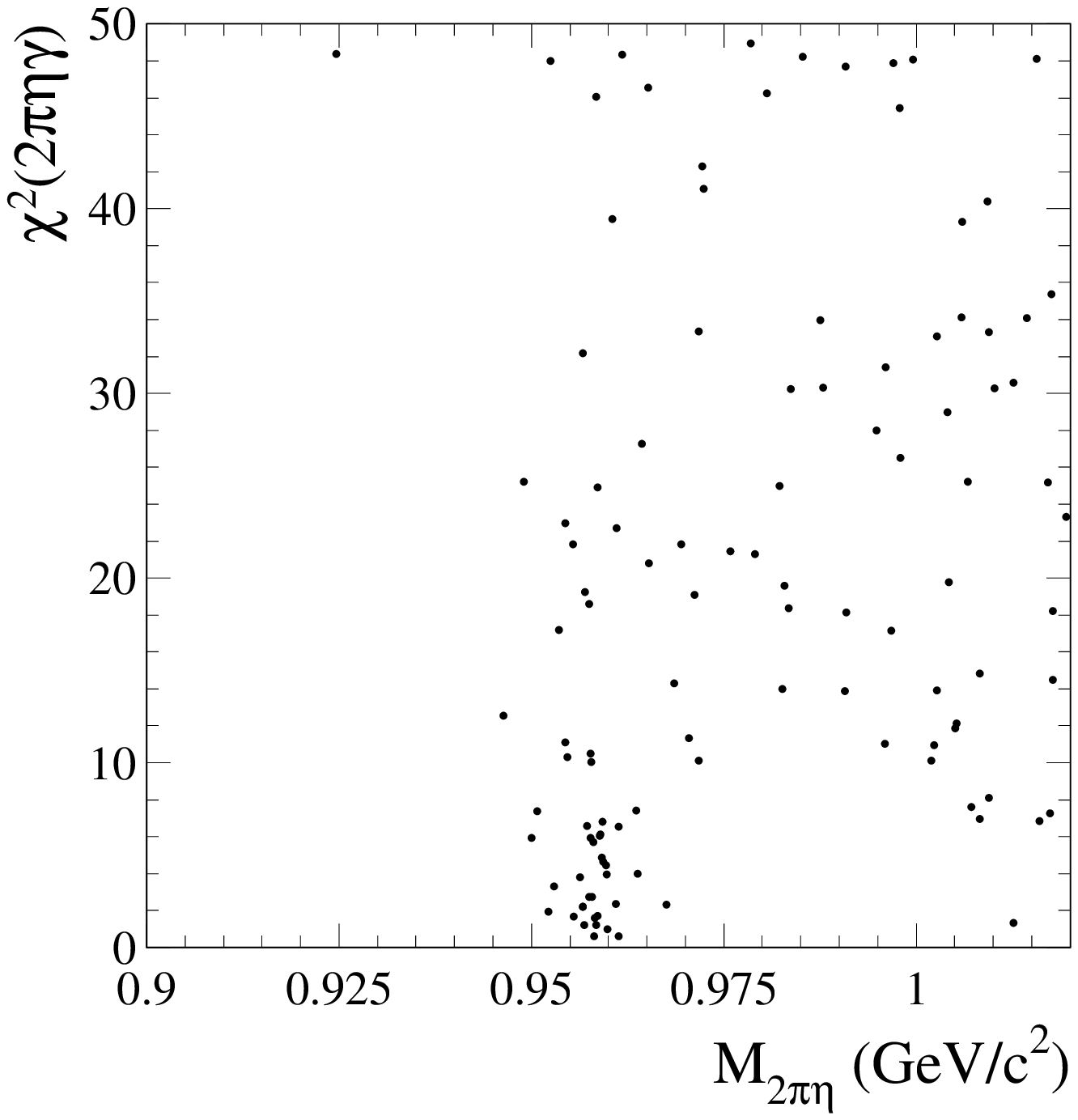}
\includegraphics[width=0.7\linewidth]{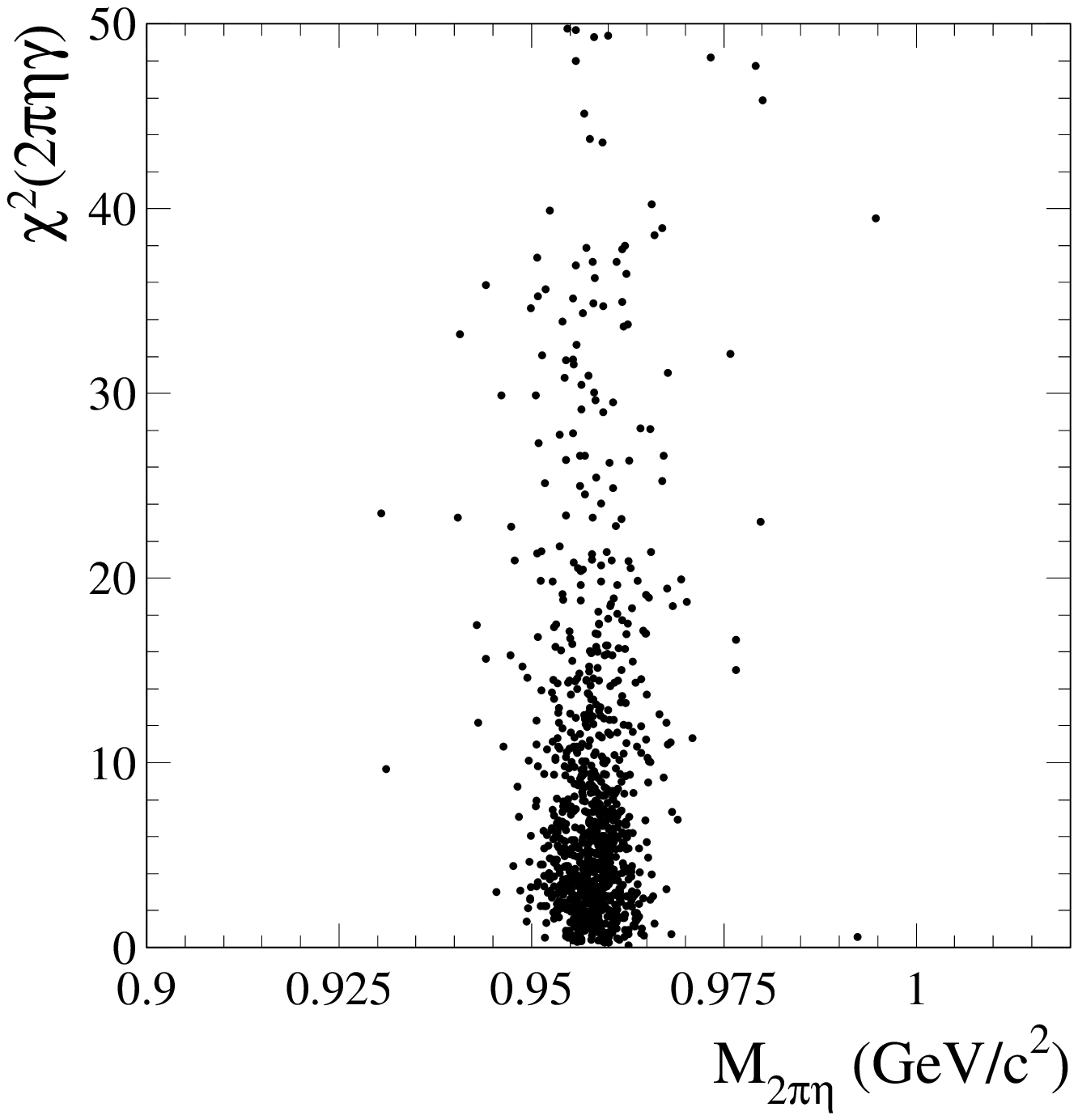}
\vspace{-0.2cm}
\caption{
  Scatter plots of \chippeg vs.\ $M_{2\pi\eta}$ for
  the selected events in the data (top) and 
  $\eetoepg \!\!\to\! \pip\pim\eta\gamma \!\!\to\! \pip\pim\gamma\gamma\gamma$ 
  signal simulation (bottom).
  }
\label{chimp}
\end{figure}

\begin{figure}
\includegraphics[width=0.8\linewidth]{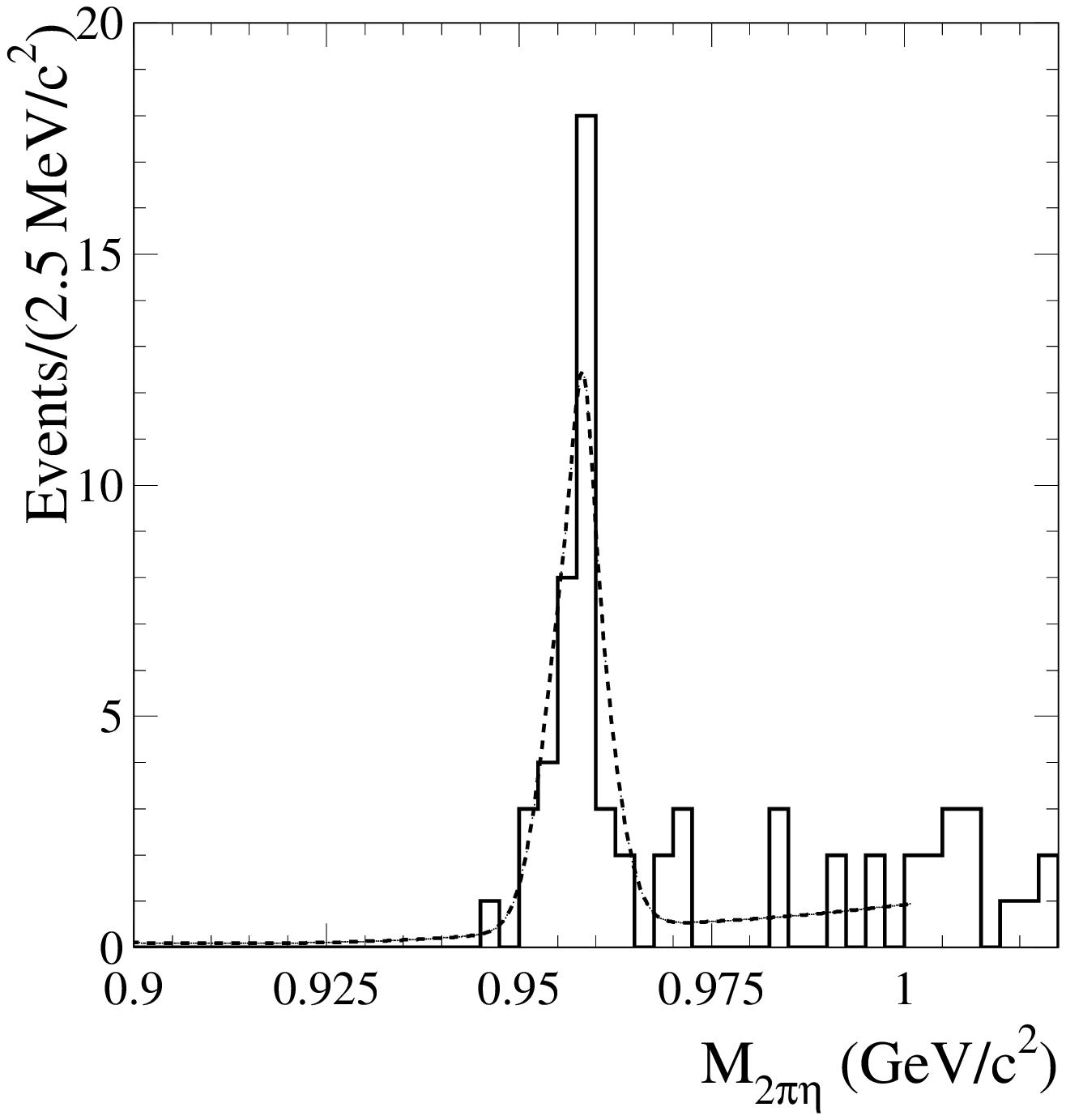}
\vspace{-0.2cm}
\caption{
  The $\pip\pim\eta$ invariant mass distribution for the 
  \eetoepg candidates in the data with
  $\eta^\prime \!\!\to\! \pip\pim\eta$, $\eta \!\!\to\! \gamma\gamma$.
  The curve represents the result of the fit described in the text.
  }
\label{etapfit1}
\end{figure}

For the \eetoepg reaction in the $\pi^+\pi^-3\gamma$ final state, 
we apply the criteria (i)--(ii) above,
the analog of (iii) $\chippeg \!\! -\! \chi^2_{2\pi 3\gamma}\! <\,$5,
and a slightly different requirement on the two-track mass of
$M_{2\pi}\! <\,$410~\mevcc, the kinematic limit for the two pions from
an $\eta^\prime \!\!\to\! \pi^+\pi^-\eta$ decay.
Figure~\ref{chimp} shows scatter plots of
\chippeg versus $M_{2\pi\eta}$ for selected
candidates in the data and the \eetoepg signal simulation;
a cluster of data events is evident near the $\eta^\prime$ mass at
small values of \chippeg.
We show the $M_{2\pi\eta}$ spectrum for data events with 
$\chippeg \! <\,$20 in Fig.~\ref{etapfit1}, 
and determine the number of events containing an $\eta^\prime$ with 
a fit to this spectrum similar to that used for the $\eta$ signal, but
over the range 900--1000~\mevcc.
The line on Fig.~\ref{etapfit1} represents the result of the fit and
the fitted number of events is 
$N_{\eta^\prime}\! =\! 38.1^{+6.8}_{-6.2}\pm1.0$.

\begin{figure}
\includegraphics[width=0.7\linewidth]{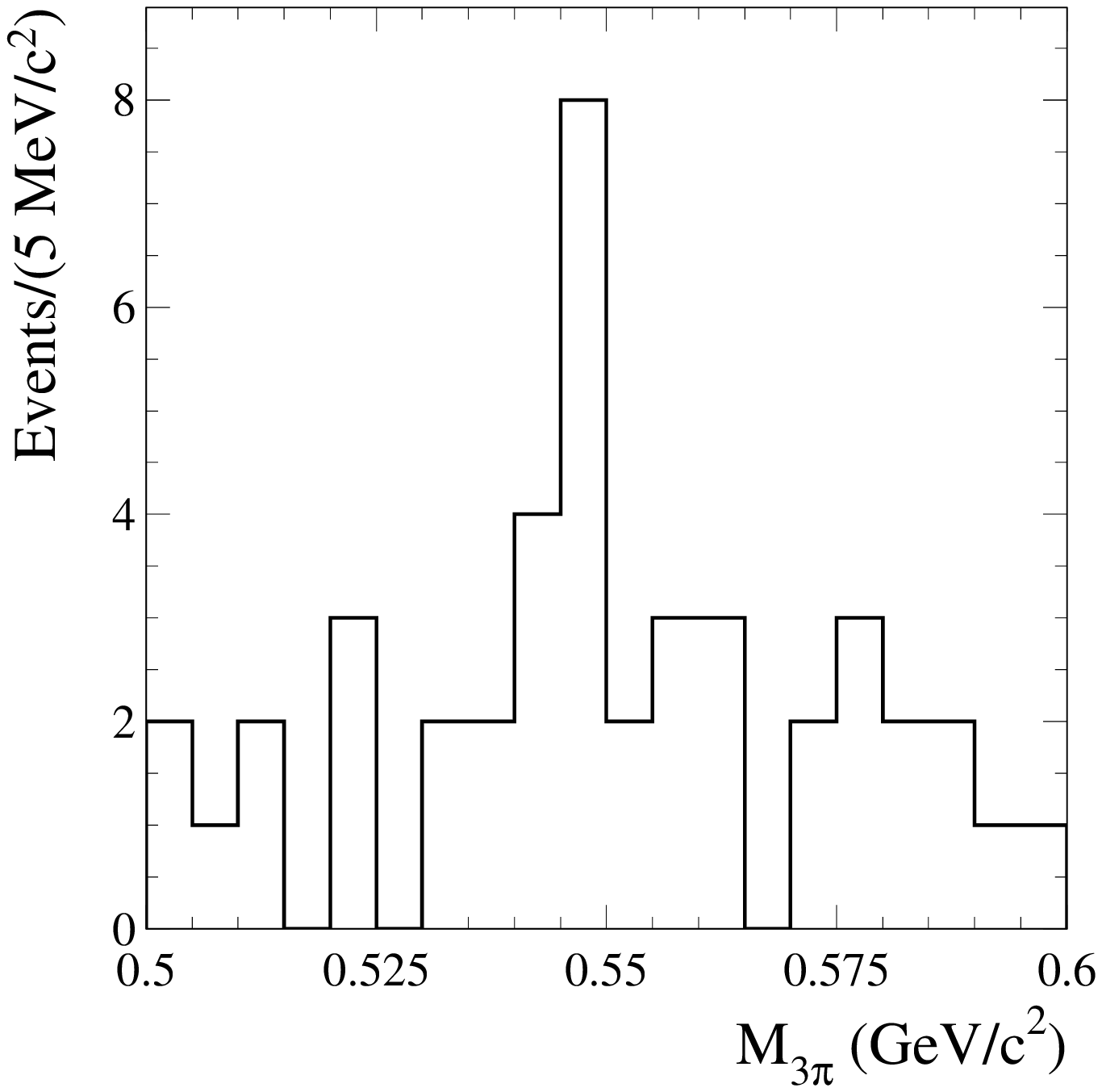}
\includegraphics[width=0.7\linewidth]{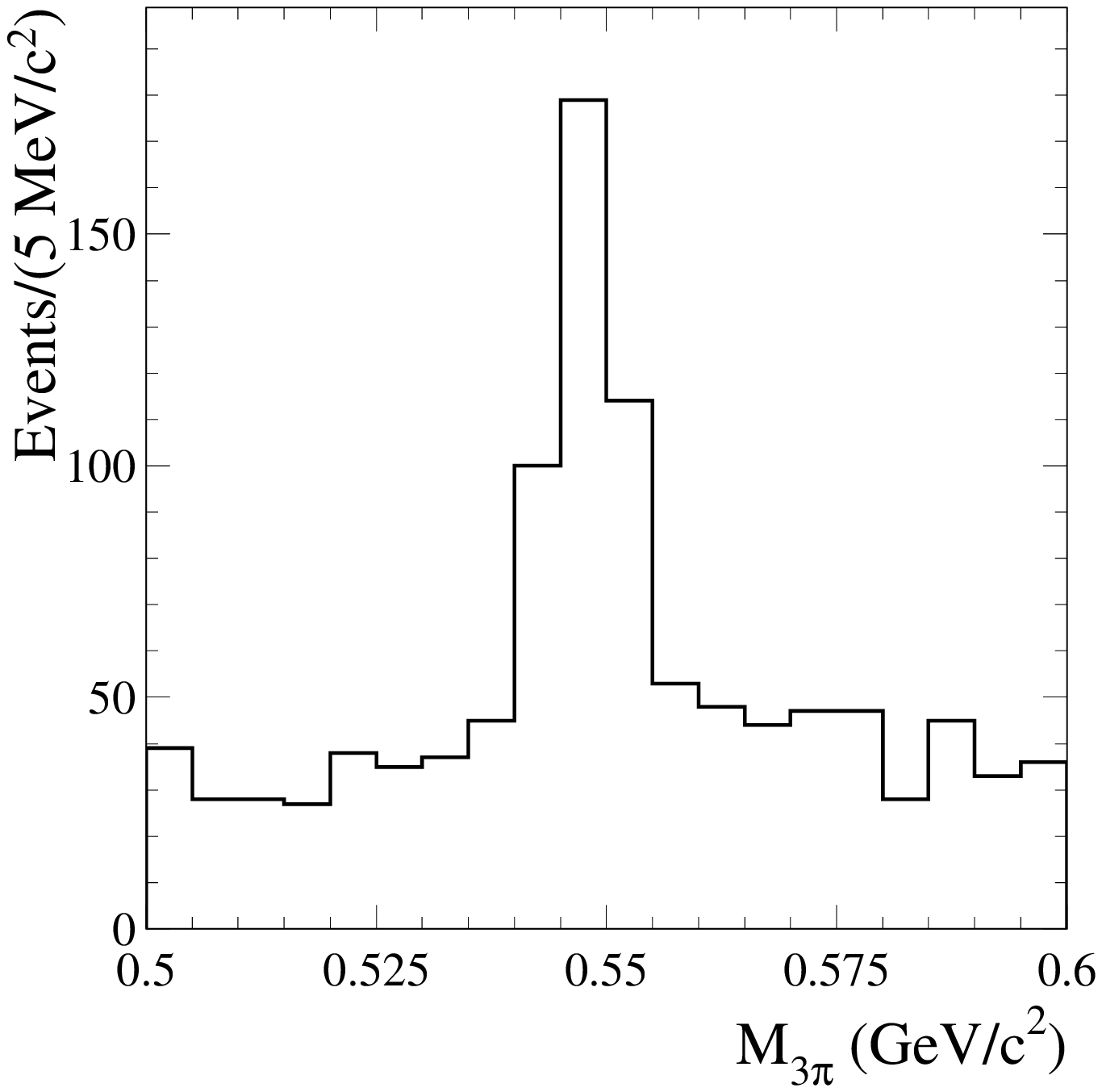}
\vspace{-0.2cm}
\caption{
  Distributions of $\pi^+\pi^-\pi^0$ invariant mass (4 combinations 
  per event) for selected events in the data (top) and 
  $\eetoepg \!\!\to\! \pip\pim\eta\gamma \!\!\to\! \pip\pim\pip\pim\pi^0\gamma$
  simulation (bottom).
\label{eta_5pi}}
\end{figure}

\begin{figure}
\includegraphics[width=0.8\linewidth]{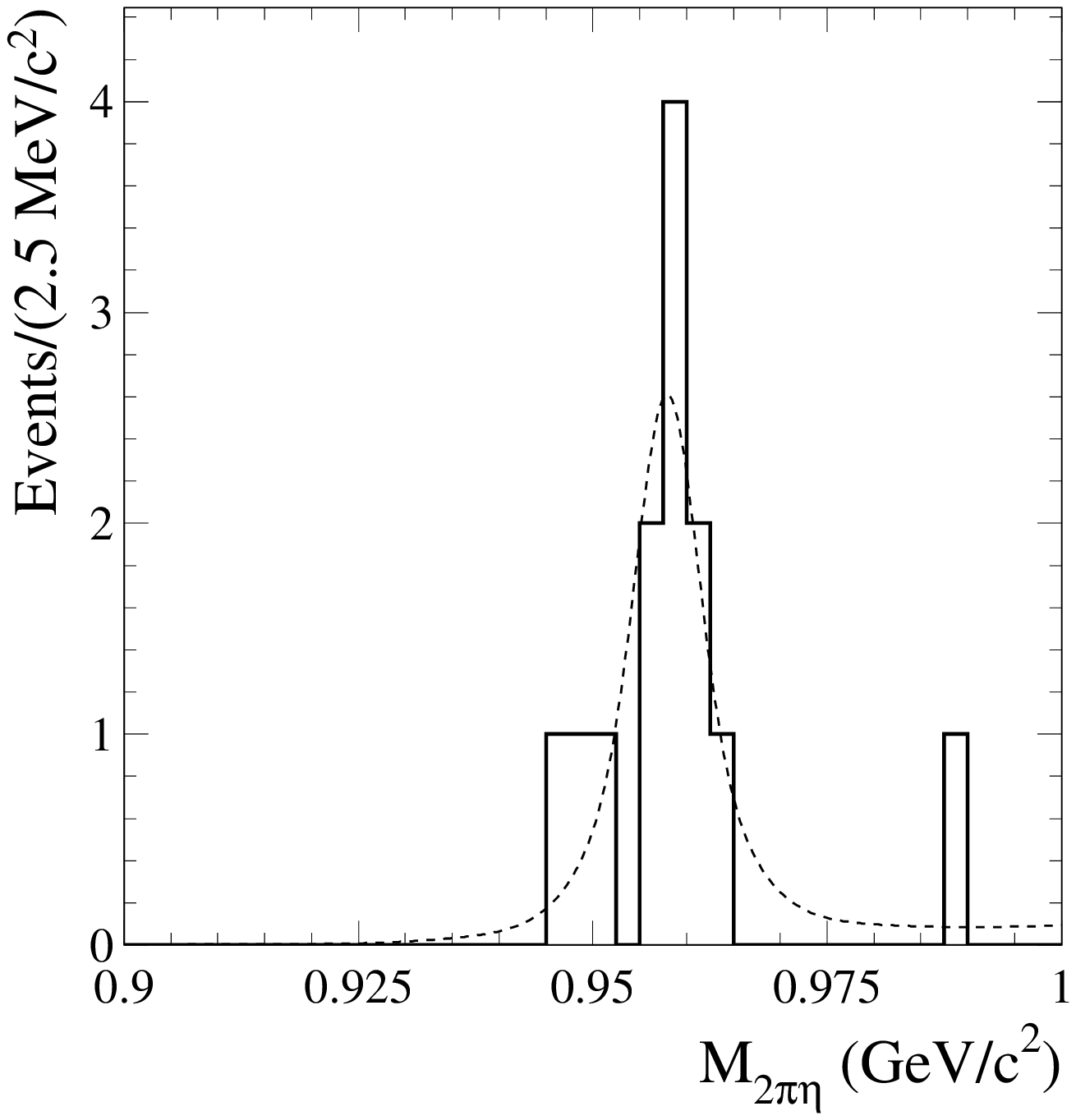}
\vspace{-0.2cm}
\caption{
  The $\pip\pim\eta$ invariant mass distribution for the 
  \eetoepg candidates in the data with
  $\eta^\prime \!\!\to\! \pip\pim\eta$, $\eta \!\!\to\! \pip\pim\pi^0$.
  The curve represents the result of the fit described in the text.
\label{etap_5pi}}
\end{figure}

For the \eetoepg reaction in the $4\pi\pi^0\gamma$ final state we require 
$\chifppng\! <\! 25$ and that none of the four charged tracks is 
identified as a kaon. 
We then search for events in which three of the pions are consistent
with an $\eta$ decay.
Figure~\ref{eta_5pi} shows the distribution of the $\pi^+\pi^-\pi^0$ 
invariant mass (4 combinations per event) for selected candidates in the
data and \eetoepg signal simulation, with the additional requirement
that $M_{5\pi}\! <1$~\gevcc.
Peaks at the $\eta$ mass are evident over a modest combinatorial background.
We select events with at least one combination in
the range 0.535$<M_{3\pi}<$0.56~\gevcc;
no event in the data or simulation has more than one.
We fit the $5\pi$ invariant mass spectrum for the selected data events 
as for the other modes, 
over the range 900--1000~\mevcc,
and show the distribution and fit result in Fig.~\ref{etap_5pi}. 
The fitted number of events containing a true $\eta^\prime$ is 
$12.0^{+3.9}_{-3.4}\pm0.3$.

\section{Background}
\label{bkgsub}
\subsection{\boldmath$e^+e^-\to\eta\gamma$}

We consider both non-peaking and peaking backgrounds, where the latter
arise from other processes producing true $\eta$ mesons or other
mesons whose decays reflect or feed down into the $\eta$ mass region.
Figure~\ref{etafit} shows that the non-peaking background is
small in the $\eta$ mass region, but increases sharply toward the
upper edge of the plot.
This is due primarily to the low-mass tail of the $\omega$ resonance in the
ISR processes $e^+e^- \!\!\to\! \gisr\pi^+\pi^-\pi^0$, and
$e^+e^- \!\!\to\! \gisr\pi^+\pi^-\pi^0\pi^0$.
Our simulation of these processes is tuned to existing data~\cite{pdg},
and predicts an $M_{3\pi}$ spectrum consistent with our selected data
both inside (excluding the $\eta$ peak) and
outside the range of Fig.~\ref{etafit}.
The simulated contributions of other ISR and $e^+e^- \!\!\to\! q\bar{q}$ 
processes to the non-peaking background are negligible.

The primary source of peaking background is the set of ISR processes
$\epem \!\!\to\! \gisr\eta\gamma$,
where the $\eta\gamma$ comes from a $\rho$, $\omega$, $\phi$,
or $J/\psi$ decay, all of which have been measured.
We calculate the number of background events using a simulation
based on the vector meson dominance model that includes
$\rho$, $\omega$, and $\phi$ amplitudes with PDG resonance 
parameters~\cite{pdg}
and phases of 0$^\circ$, 0$^\circ$, and 180$^\circ$, respectively, and
describes the existing data on the $\epem \!\!\to\! \eta\gamma$
reaction in the $\rho$-$\omega$-$\phi$ mass region~\cite{SND,CMD2}.
The model also includes $J/\psi$ production, and
predicts a total peaking background of 2.6$\pm$0.5 events. 

\begin{figure}
\includegraphics[width=0.7\linewidth]{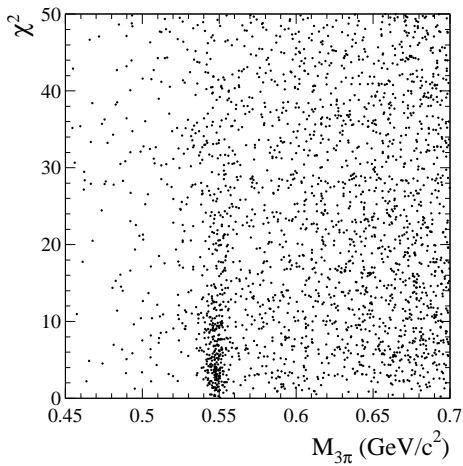}
\vspace{-0.2cm}
\caption{
  Scatter plot of $\chi^2_{3\pi\gamma\gamma}$ versus $M_{3\pi}$ for
  the selected data events containing an additional photon.
  }
\label{etaggsp}
\end{figure}

\begin{figure}
\includegraphics[width=0.7\linewidth]{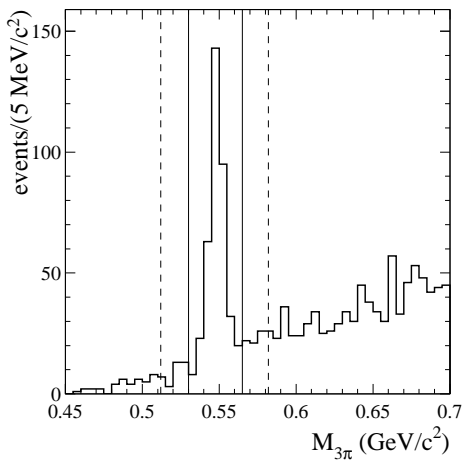}
\vspace{-0.2cm}
\caption{
  The $3\pi$ invariant mass spectrum for events in the data with
  $\chi^2_{3\pi\gamma\gamma} < 25$.  
  The solid vertical lines bound the $\eta$ signal region;
  the sideband regions are between these and the dashed lines.}
\label{etgg_m3pi}
\end{figure}

The simulation does not include other contributions such as decays of excited
$\rho$, $\omega$, or $\phi$ states, as they are unmeasured and
expected to be small.
As a check, we select $\epem \!\!\to\! \gisr\eta\gamma$ events explicitly
from our data, by subjecting any event with an additional photon to a
kinematic fit to the $3\pi\gamma\gamma$ hypothesis.
Figure~\ref{etaggsp} shows a scatter plot of the $\chi^2$ of this fit 
($\chi^2_{3\pi\gamma\gamma}$) 
versus the $3\pi$ invariant mass, and Fig.~\ref{etgg_m3pi} shows
the $M_{3\pi}$ spectrum for events with 
$\chi^2_{3\pi\gamma\gamma} \! <\,$25; 
a strong $\eta$ signal is present.
We estimate the number of $\epem \!\!\to\! \gisr\eta\gamma$ events by
counting the events in the signal region indicated in Fig.~\ref{etgg_m3pi} 
and subtracting the number in the two sidebands. 
The resulting number of events, 274$\pm$22, is consistent with the
261$\pm$5$\pm$9 expected from the simulation, where the
systematic error in the latter is due to experimental uncertainties 
on the input parameters to the simulation. 
Repeating this exercise in several different ranges of the 
$\eta\gamma$ invariant mass, we obtain the
results listed in Table~\ref{etggtab};
data and simulation are consistent.

\begin{table}
\caption{
  The number of selected $\epem \!\!\to\! \gisr\eta\gamma$ 
  events in the data in several ranges of the $\eta\gamma$ invariant
  mass compared with expectations from the simulation.
  The first error on each expected number is statistical, the second
  systematic.
  }
\begin{center}
\begin{tabular}{|c|r@{$\pm$}l|r@{$\pm$}l@{$\pm$}l|} \hline
$M_{\eta\gamma}$ (GeV/$c^2)$ & \multicolumn{2}{c|}{$N_{data}$}
                             & \multicolumn{3}{c|}{$N_{expect}$} \\
\hline
0.55--0.95 &  25 &   9 &  43 &   3 & 4\\
0.95--1.05 & 200 &  15 & 192 &   5 & 4\\
1.05--3.05 &  18 &  12 &   5 &   1 & 6\\
3.05--3.15 &  31 &   6 &  21 &   1 & 2\\
3.15--6.50 & 0.0 & 1.4 &   1 & 0.4 & 2\\
\hline
0.55--6.50 & 274 &  22 & 261 &   5 & 9\\
\hline
\end{tabular}
\label{etggtab}
\end{center}
\end{table}

Other possible sources of peaking background are the processes
$\epem \!\!\to\! VP \!\!\to\! \eta\pi^0\gamma$, where $V$ denotes a
vector meson, $\rho$, $\omega$, or $\phi$, and $P$ is a $\pi^0$ or $\eta$.
The CLEO and BES experiments have measured these cross sections
at $\sqrt{q^2}\approx\,$3.7~\gev~\cite{CLEOVP,BESVP};
assuming the $1/q^4$ dependence of $VP$ form factors predicted by perturbative 
QCD~\cite{chernyakPR}, we estimate the $\epem \!\!\to\! \eta\pi^0\gamma$
cross section to be about 3~fb at our c.m.\ energy.
The simulated selection efficiency is very low due to the additional $\pi^0$,
approximately 2$\times\! 10^{-4}$, so we expect only 0.2 background
events from this source.
The corresponding ISR process $\epem \!\!\to\! \gisr VP$ can also contribute, 
and we estimate a cross section of about 13~fb, 
(for 20$^\circ\! <\!\theta^\ast_\gamma\!<$160$^\circ$),
based on our studies of several ISR final states with $VP$
components~\cite{BAD767,isrrefs}, 
including $4\pi$, $3\pi$, $3\pi\eta$, $2\pi\eta$, $2K\pi^0$, and $2K\eta$.
This cross section is relatively large, one-quarter of the 
$\epem \!\!\to\! \gisr\eta\gamma$ cross section, but the selection
efficiency is less than 2$\times\! 10^{-5}$,
so we expect no more than 0.1 events from this source.

The $\epem \!\!\to\! \gisr\eta\pi^0\gamma$ and 
$\epem \!\!\to\! \eta\pi^0\gamma$ events are selected
about 100 times more efficiently by the $\eta\gamma\gamma$ criteria
than by the $\eta\gamma$ criteria, and similar factors apply to
other types of events containing additional pions and/or photons.
We can therefore make another estimate of their overall contribution
from the difference between the observed and expected numbers of
$\epem \!\!\to\! \gisr\eta\gamma$ candidates of 13$\pm$24
(Table~\ref{etggtab}).
Accounting for the $\sim$10\% uncertainty in relative selection efficiencies,
we estimate $<\,$0.6 such events in our signal peak at the 90\% CL.

ISR production of an $\Upsilon(1S)$, $\Upsilon(2S)$, or $\Upsilon(3S)$
resonance could produce a peaking background if the $\Upsilon$
decays to $\eta\gamma$, since the ISR photon is rather soft.
From the upper limit on ${\cal B}(\Upsilon(1S) \!\!\to\! \eta\gamma)$
of 2.1$\times\! 10^{-5}$~\cite{pdg}, we estimate that the number of 
$\epem \!\!\to\! \gisr\Upsilon(1S) \!\!\to\! \eta\gamma\gamma$ 
events in our data does not exceed 100. 
Using the relation
$$\frac{\Gamma(\Upsilon(nS)\to\eta\gamma)}{\Gamma(\Upsilon(1S)\to\eta\gamma)}
\approx
\frac{\Gamma(\Upsilon(nS)\to e^+e^-)}{\Gamma(\Upsilon(1S)\to e^+e^-)},\;n=2,3$$
we obtain corresponding limits for $\Upsilon(2S)$ and $\Upsilon(3S)$ of
50 and 140, respectively. 
The selection efficiencies for the $1S$, $2S$, and $3S$ processes are
below 0.01\%, 0.02\%, and 0.08\%, respectively, so
the total $\Upsilon$ background does not exceed 0.13 events.

We search for peaking background in the $\epem \!\!\to\! q\overline{q}$ process
using the JETSET simulation.
From 736 million simulated events (corresponding to about twice our 
integrated luminosity) only two events pass the $\eta\gamma$
selection criteria. 
Only one of them, a $K^0\overline{K}^0\eta$ final state, has a $3\pi$
invariant mass close to the $\eta$ mass. 
Since we do not expect JETSET to predict rates for such rare events correctly,
we select $\epem \!\!\to\! K\overline{K}\eta$, $\eta \!\!\to\! \gamma\gamma$
events from our data as a check.
We perform a kinematic fit to the $K^+K^-\gamma\gamma$ hypothesis on
all events with at least one charged track identified as a kaon, 
and select events with $\chi^2_{2K\gamma\gamma}\! <$10.
From the 2$\pm$30 events found in the data and 312$\pm$14 expected
from the simulation,
we conclude that JETSET overestimates the yield and that this source
of background is negligible.

Taking the estimate of the number of peaking background events from 
$\epem \!\!\to\! \gisr\eta\gamma$, 
and considering the upper limits on all the other sources as
additional systematic errors, 
we estimate the total peaking background to be 2.6$\pm$0.8 events.
Subtracting this from the number of observed events with a true $\eta$,
we obtain the number of detected \eetoeg events:
$$N_{\eta\gamma}=20.1^{+5.6}_{-4.9}\pm1.0.$$

\subsection{\boldmath$e^+e^-\to\eta^\prime\gamma$\label{bkg_etap}}

We estimate backgrounds in the \eetoepg sample using similar procedures.
The non-peaking background is very small for both the 
$\eta \!\to\! \gamma\gamma$ (see Fig.~\ref{etapfit1}) and
$\eta \!\to\! \pi^+\pi^-\pi^0$ (see Fig.~\ref{etap_5pi}) modes.
According to the simulations, it is dominated by the ISR processes
$\epem \!\!\to\! \gisr\pi^+\pi^-\pi^0$ and
$\epem \!\!\to\! \gisr\pi^+\pi^-\pi^0\pi^0$.
As for the \eetoeg process, the simulated non-peaking background mass 
distributions are consistent with those observed in data.

The largest source of peaking background in the simulations is the ISR process
$\epem \!\!\to\! \gisr\eta^\prime\gamma$,
where the $\eta^\prime\gamma$ comes mainly from $\phi$ and $J/\psi$ decays.
These have been measured with about 10\% accuracy, and we use a 
vector-dominance based simulation similar to that for the $\eta\gamma$ analysis
to estimate their contribution.
In addition to the $\phi$ and $J/\psi$, we include contributions from
the high-mass tails of the $\rho$ and $\omega$ with couplings
determined from the measured $\eta^\prime \!\!\to\! \omega\gamma$ and 
$\rho\gamma$ decay widths. 
We estimate a peaking background from this source of 0.3$\pm$0.1
events in each of the two $\eta$ decay modes.

\begin{figure}
\begin{center}
\includegraphics[width=0.7\linewidth]{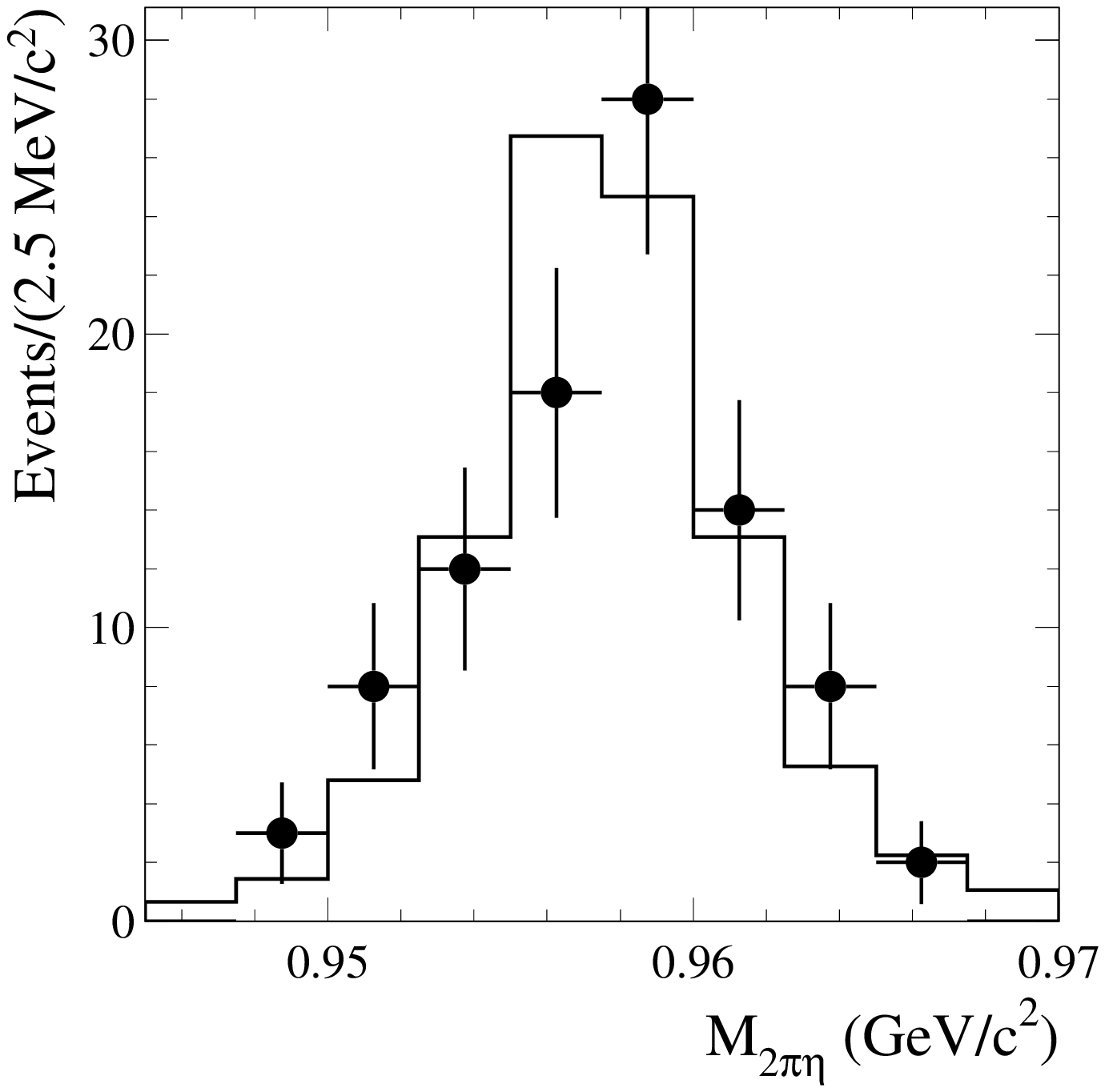}
\vspace{-0.2cm}
\caption{
  Distributions of the $\pip\pim\eta$ invariant mass for selected 
  $\epem \!\!\to\! \gisr J/\psi \!\!\to\! \gisr\eta^\prime\gamma$
  events in the data (points with error bars) and simulation (histogram).
\label{metapg}}
\end{center}
\end{figure}

We check this prediction by selecting
$\epem \!\!\to\! \gisr\eta^\prime\gamma$ events using a kinematic
fit to the $\pi^+\pi^-\eta\gamma\gamma$ hypothesis. 
Selecting events with a $\chi^2_{2\pi\eta\gamma\gamma}\! <\! 25$, we count
signal and sideband events in the $\pi^+\pi^-\eta$ invariant mass
distributions to obtain numbers of events from this source 
in a set of $\eta^\prime\gamma$ mass intervals.
The results from data and simulation listed in Table~\ref{etpggtab}
are consistent.
In the $J/\psi$ mass region these events are practically free of background, 
and we compare the data and simulated $\pi^+\pi^-\eta$ invariant mass
distributions for events with an $\eta^\prime\gamma$ mass in the range
3.05--3.15~\gevcc in Fig.~\ref{metapg}. 
The RMS of the distribution is 3.9$\pm$0.3~\mevcc in the data and 
3.80$\pm$0.06~\mevcc in the simulation.

\begin{table}
\caption{
  The number of selected $\epem \!\!\to\! \gisr\eta^\prime\gamma$
  events in the data in several ranges of the $\eta^\prime\gamma$ invariant
  mass compared with expectations from simulation.
  The first error on each expected number is statistical, the second
  systematic.
  }
\begin{center}
\begin{tabular}{|c|r@{$\pm$}l|r@{$\pm$}l@{$\pm$}l|} \hline
$M_{\eta^\prime\gamma}$ (GeV/$c^2)$ & \multicolumn{2}{c|}{$N_{data}$}
& \multicolumn{3}{c|}{$N_{expect}$} \\
\hline
$< $1.5    & $-$2& 12  & 1.7 & 0.3 & 0.2\\
1.5--2.0   &  6  &   4 & 1.0 & 0.2 & 1.0\\
2.0--3.0   &  2  &   3 & 1.3 & 0.3 & 1.3\\
3.0--3.2   & 97  &  10 & 102 &  2  & 10\\
$>$  3.2   &  3  &   3 & 1.1 & 0.2 & 1.1\\
\hline
Total & 110 &  17 & 107 &   2 & 10\\
\hline
\end{tabular}
\label{etpggtab}
\end{center}
\end{table}

To bound peaking background from $\epem \!\!\to\! \eta^\prime\pi^0\gamma$,
$\epem \!\!\to\! \gisr\eta^\prime\pi^0\gamma$ and other events containing
additional pions and/or photons, we consider the difference between the
observed and expected numbers of $\epem \!\!\to\! \gisr\eta^\prime\gamma$ 
candidates in Table~\ref{etpggtab}, 3$\pm$20.
Taking into account the factor of 30 difference in selection
efficiency, along with its 10\% systematic uncertainty,
we determine that the total background contribution from such processes
does not exceed 1 event in the $\eta \!\!\to\! \gamma\gamma$ mode or
0.3 events in the $\eta \!\!\to\! \pi^+\pi^-\pi^0$ mode.

From the upper limit
$\BR (\Upsilon(1S) \!\!\to\! \eta^\prime\gamma)\! 
<\! 1.6\times 10^{-5}$~\cite{pdg}, we estimate that the number of
$\epem \!\!\to\! \gisr\Upsilon \!\!\to\! \gisr\eta^\prime\gamma$
events in our data does not exceed 80, 40, or 110 for the $1S$, $2S$, or $3S$
states, respectively.
The simulated efficiencies for such events to pass the \eetoepg
selection criteria are small, and we estimate that the total 
$\Upsilon$ background does not exceed 0.03 events for the 
$\eta \!\!\to\! \gamma\gamma$ mode, and is negligible for the
$\eta \!\!\to\! \pi^+\pi^-\pi^0$ mode.
In the 736 million $\epem \!\!\to\! q\overline{q}$ events simulated
by JETSET, we find none that passes the $\eta^\prime\gamma$ 
selection criteria.

Considering the upper limits as systematic errors, we estimate total peaking
backgrounds of 0.3$\pm$1.0 and 0.3$\pm$0.3 events in the $2\pi 3\gamma$
and $4\pi 3\gamma$ final states, respectively.
Subtracting these from the numbers of observed $\eta^\prime$ events, we obtain
a total number of $\eta^\prime\gamma$ events,
$$N_{\eta^\prime\gamma}=49.5^{+7.7}_{-7.1}\pm1.5.$$

\section{Detection efficiency}
\label{deteff}
\subsection{\boldmath $e^+e^-\to \eta\gamma$\label{detef_eta}}

The detection efficiency determined from the simulation is
$\varepsilon_{MC}\! =\! (2.01\pm0.06)\%$, 
where the error includes a statistical error and the uncertainty in
the value of $\BR (\eta \!\to\! \pi^+\pi^-\pi^0)$.
This efficiency must be corrected to account for deficiencies in the simulated
detector response.
We take advantage of the relatively large cross section for the ISR process
$\epem \!\!\to\! \gisr\omega(782) \!\!\to\! \gisr\pi^+\pi^-\pi^0$, which
can be selected with very low background~\cite{BAD767}. 
The $M_{3\pi}$ spectrum for this process is described by
\begin{equation}
\frac{{\rm d}N}{{\rm d}M} =
\sigma_{3\pi}(M)\frac{{\rm d}L}{{\rm d}M}\,R\,\varepsilon (M),
\label{thmspt}
\end{equation}
where $\sigma_{3\pi}(M)$ is the Born cross section for $\epem \!\!\to\! 3\pi$,
${{\rm d}L}/{{\rm dM}}$ is the so-called ISR differential luminosity, 
$\varepsilon (M)$ is the detection efficiency as a function of mass,
and $R$ is a radiative correction factor (see Ref.~\onlinecite{BAD767} for
a more detailed discussion).
The $\epem \!\!\to\! 3\pi$ Born cross section near the $\omega$ mass
can be described by a Breit-Wigner function with well measured 
parameters~\cite{pdg}.
We calculate the ISR luminosity from the total integrated luminosity
$L$ and the theoretical ISR photon radiator function~\cite{ivanch}. 
The radiative correction factor is known with a theoretical 
uncertainty below 1\%~\cite{strfun}.
We can therefore fit the $3\pi$ invariant mass spectrum for events passing the
criteria for this analysis in the $\omega$ mass region to determine the
efficiency directly from the data.

\begin{figure}
\includegraphics[width=0.7\linewidth]{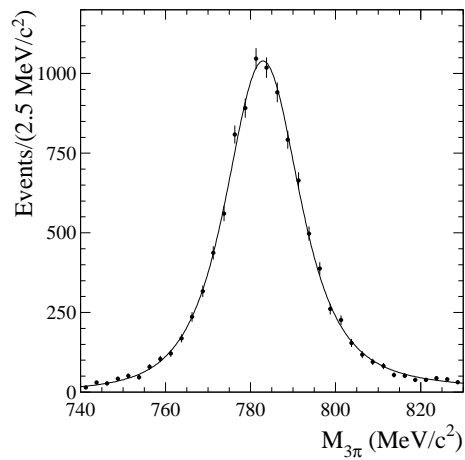}
\vspace{-0.2cm}
\caption{
  The $\pip\pim\pi^0$ invariant mass spectrum for data events in the $\omega$ 
  mass region. The curve is the result of the fit described in the
  text.
  }
\label{fitomega}
\end{figure}

Figure~\ref{fitomega} shows this distribution after subtraction of
the $\sim$0.5\% background, estimated from simulation
as described in Ref.~\onlinecite{BAD767}.
The fitting function is given by Eq.(\ref{thmspt})
convolved with the simulated detector resolution function.
There are three free parameters: 
the efficiency correction factor 
$\delta_\omega$ ($\varepsilon\! =\!\delta_\omega \varepsilon_{MC}$);
the $\omega$ mass;
and $\sigma_G$, an ad-hoc Gaussian smearing to account for any
resolution difference between data and simulation. 
The curve in Fig.~\ref{fitomega} represents the result of the fit, which
returns:
\begin{eqnarray}
\delta_\omega & = & 0.933\pm0.009\pm0.026,\nonumber \\
   \sigma_G   & = & 1.3^{+0.5}_{-1.0}\pm0.3 \mevcc,
\end{eqnarray}
where the first error is statistical and the second systematic. 
The fitted mass is shifted from the nominal value by 0.5~\mevcc,
consistent with expectations from our detector simulation.
The systematic error in the correction factor includes contributions from
simulation statistics (1.2\%), 
uncertainties on the radiative correction (1\%), 
background subtraction (0.2\%),
and the PDG $\omega$ width (1.5\%) and peak cross section (1.5\%).
The systematic error in $\sigma_G$ is due to the uncertainty
in the $\omega$ width.

\begin{figure*}
\includegraphics[width=0.325\linewidth]{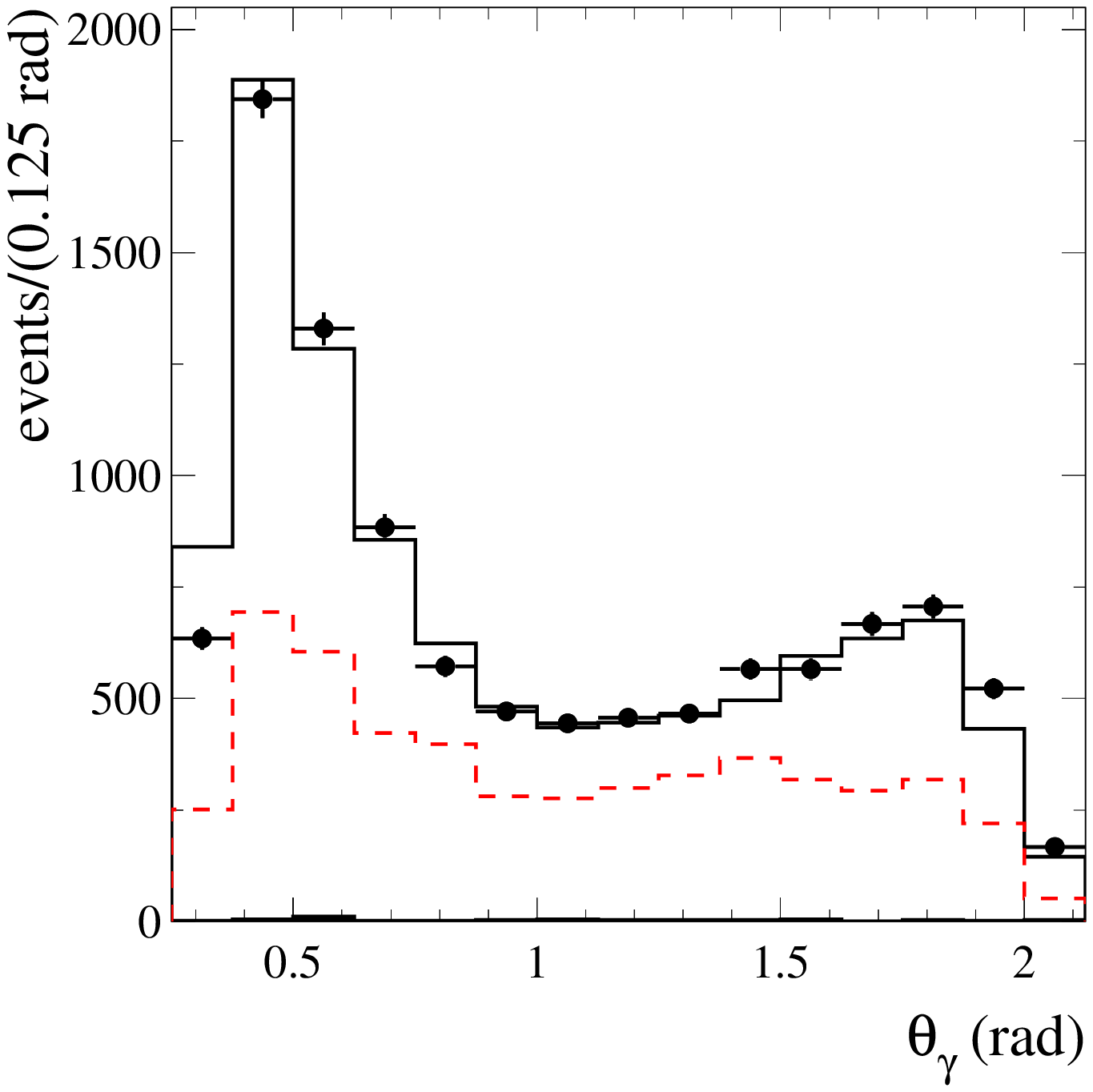}
\includegraphics[width=0.325\linewidth]{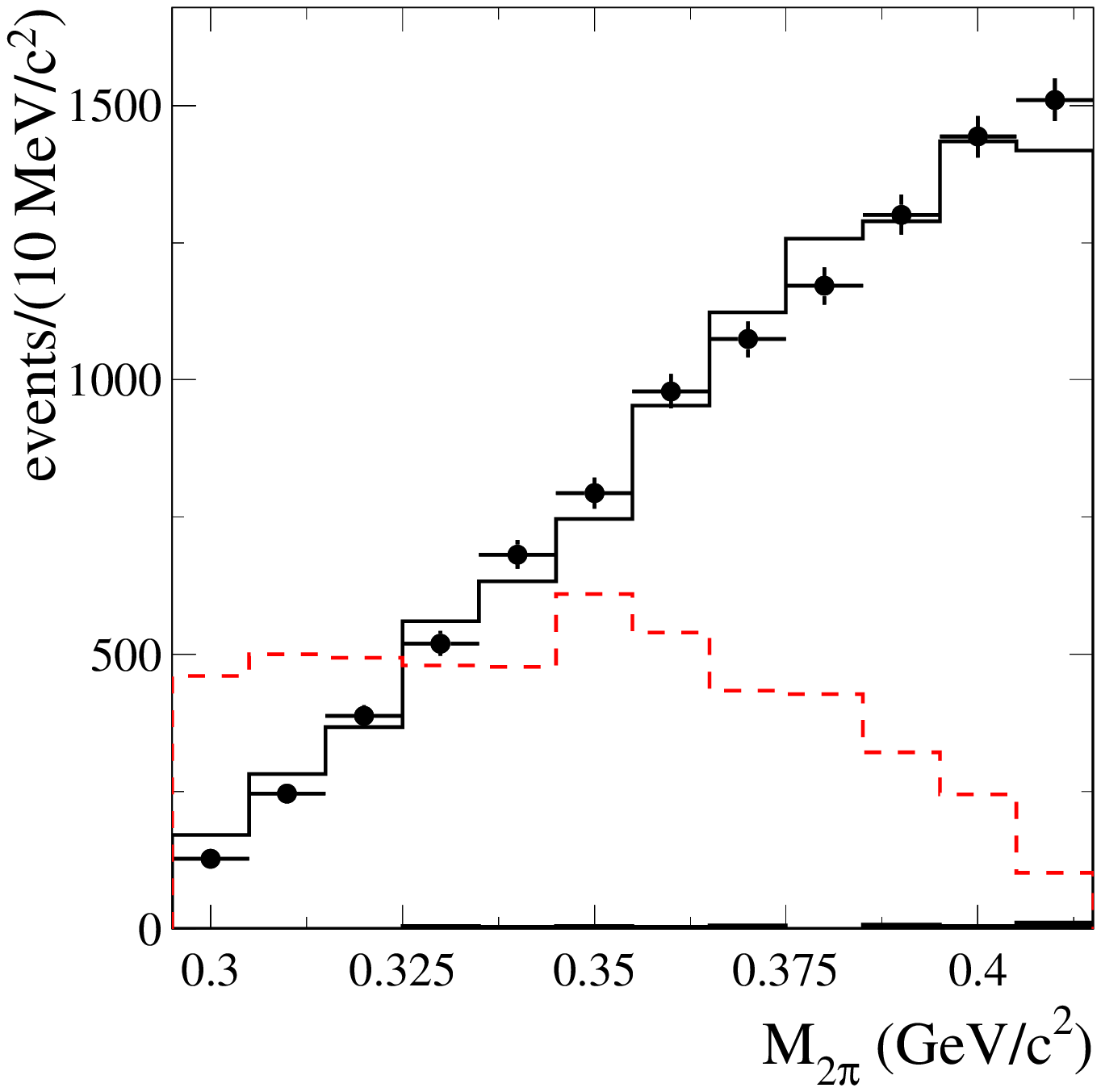}
\includegraphics[width=0.325\linewidth]{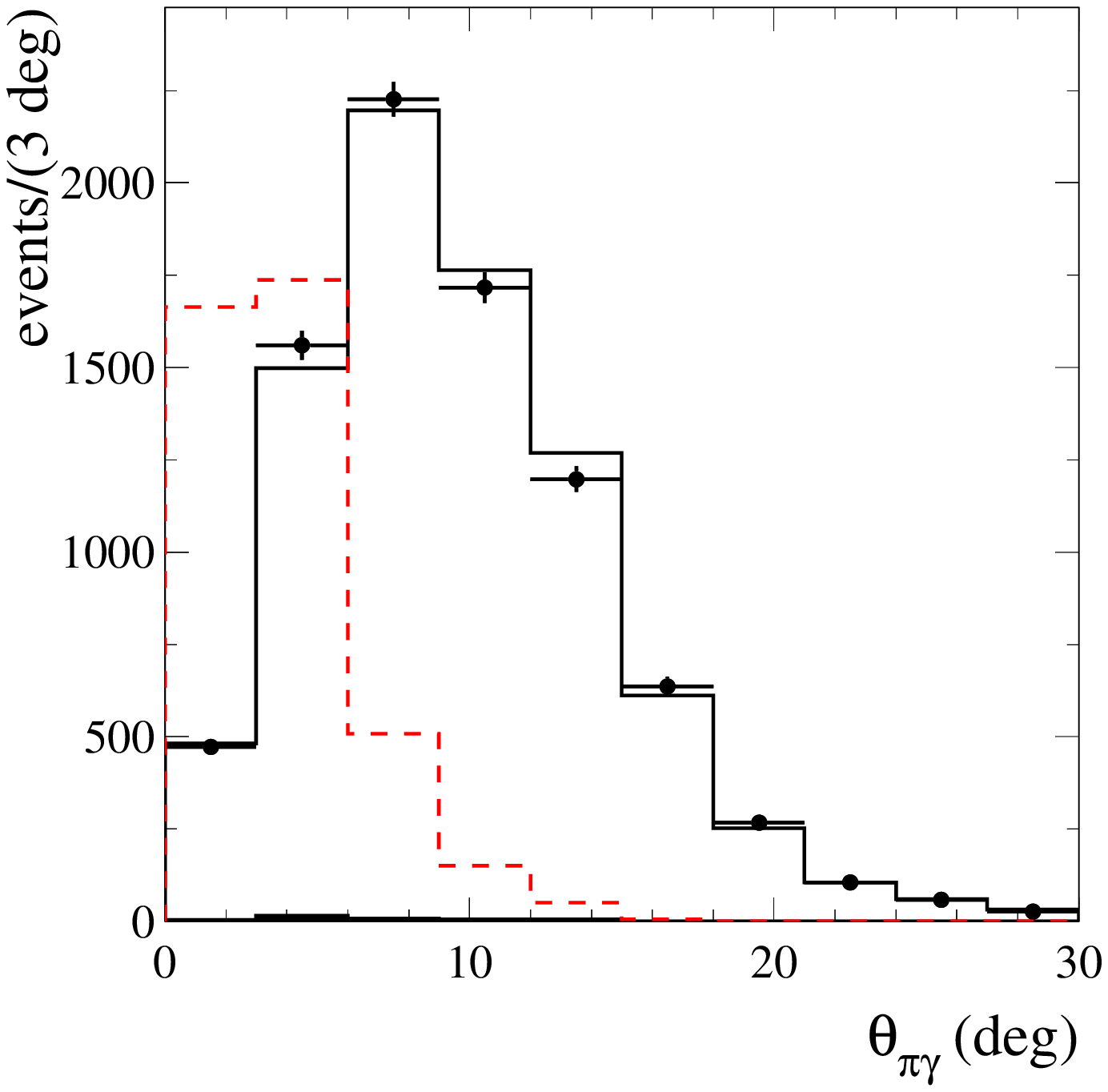}
\vspace{-0.2cm}
\caption{
  Distributions of the photon polar angle (left),
  the invariant mass of the two charged pions (middle), and
  the minimum angle between a charged pion and a photon from the
  $\pi^0$ decay (right) for data (points with error bars) and
  simulated (solid lines) $\epem \!\!\to\! \gisr\omega$ events.
  The simulated background is shown as the (very small) shaded histograms,
  and the dashed lines show the distributions for simulated \eetoeg events. 
  }
\label{effcor}
\end{figure*}

\begin{figure*}
\includegraphics[width=0.325\linewidth]{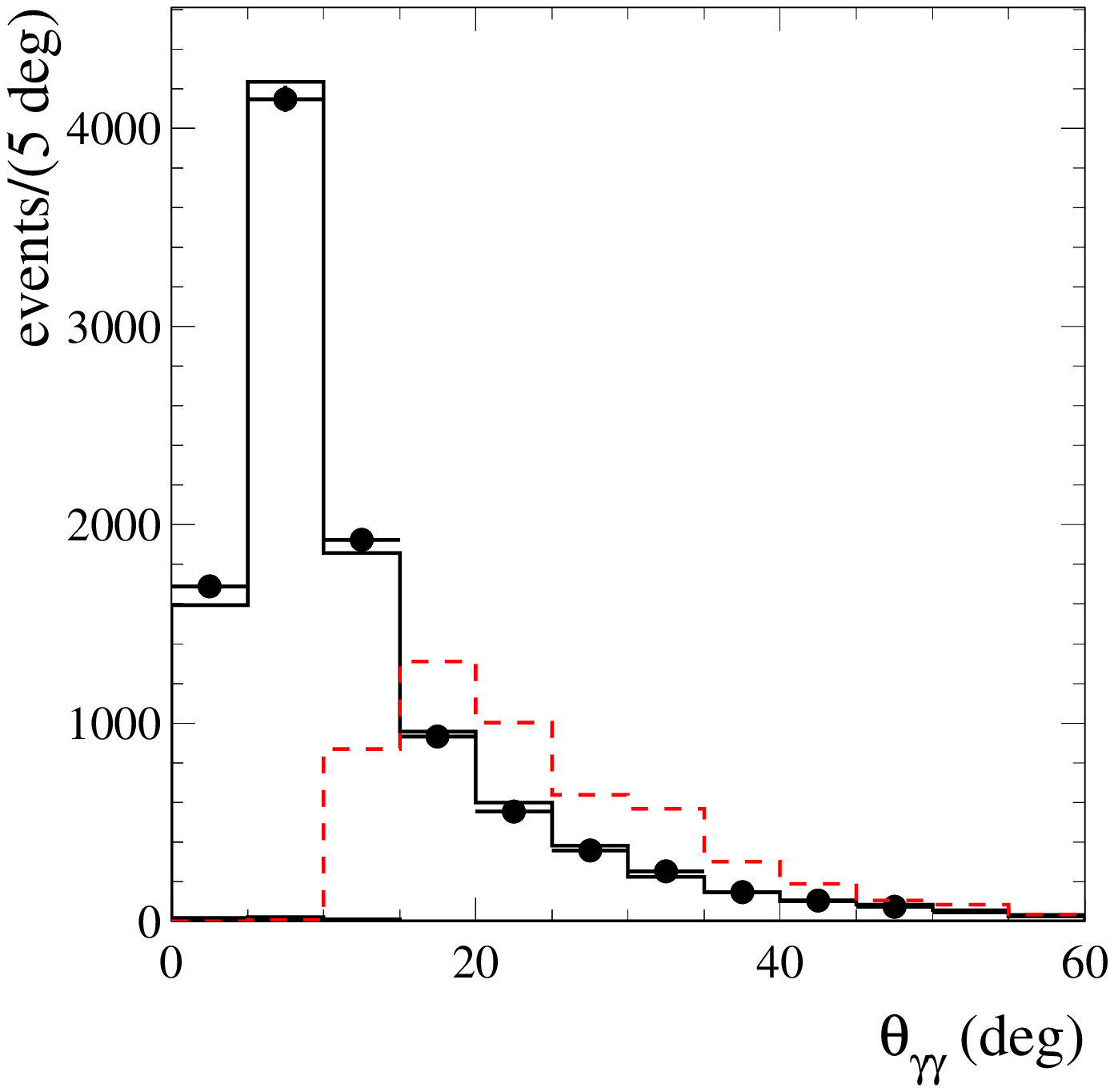}
\includegraphics[width=0.325\linewidth]{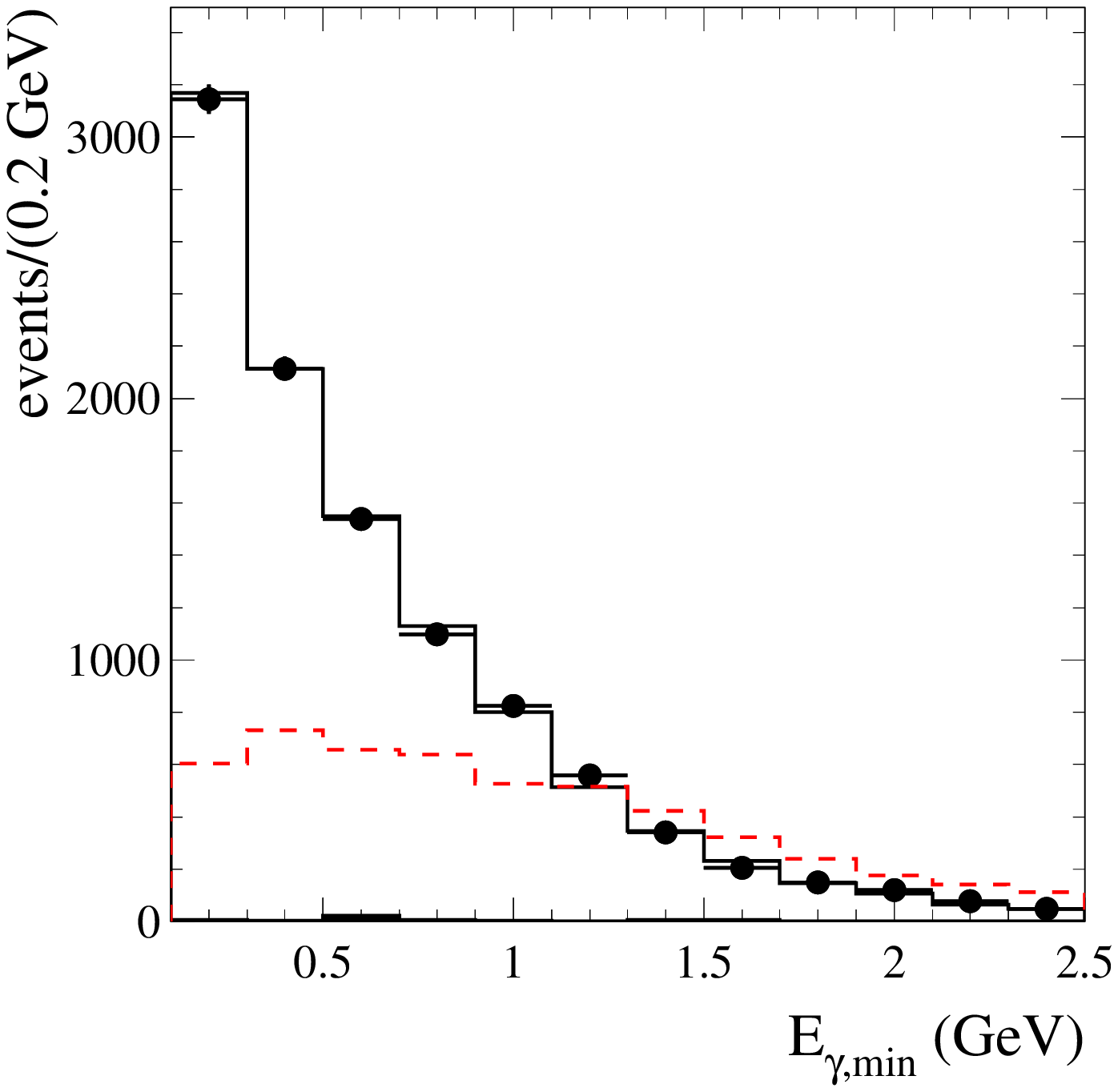}
\includegraphics[width=0.325\linewidth]{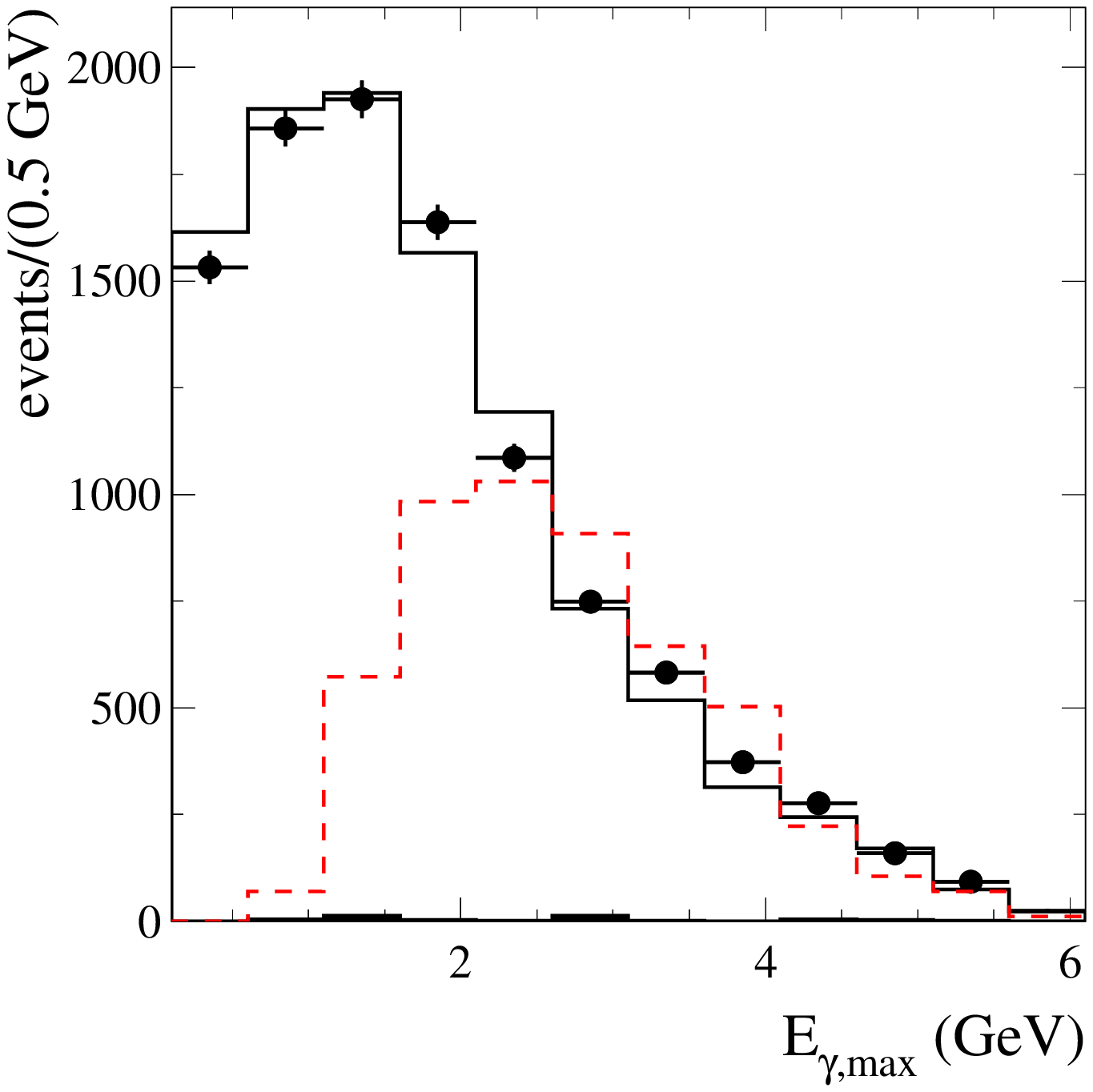}
\vspace{-0.2cm}
\caption{
  Distributions of the angle between the two photons from the 
  $\pi^0 \!\!\to\! \gamma\gamma$ or $\eta \!\to\! \gamma\gamma$ decay (left),
  and the minimum (middle) and maximum (right) energy of the decay photons,
  for data (points with error bars) and simulated (solid lines)
  $\epem \!\!\to\! \gisr\omega$ events,
  simulated background events (small shaded histograms), and
  simulated $\eetoepg \!\!\to\! \pi^+\pi^-\eta$ events (dashed lines).
\label{effcor_etap}}
\end{figure*}

Before applying this correction to the \eetoeg efficiency,
we must take into account differences in the distributions of any
kinematic variables on which the efficiency depends.
Of the many variables studied, three show large differences,
the photon polar angle $\theta_\gamma$,
the invariant mass of the two charged pions $M_{2\pi}$,
and the minimum angle between a charged pion and a photon from the
$\pi^0$ decay $\theta_{\pi\gamma}$;
in Fig.~\ref{effcor} we compare their distributions in simulated 
$\epem \!\!\to\! \gisr\omega$ events (solid histograms) with those in 
simulated \eetoeg events (dashed histograms).
The $\theta_{\pi\gamma}$ distribution for the $\epem \!\!\to\! \gisr\omega$
data (dots in Fig.~\ref{effcor}) is consistent with that for the
simulation,
but significant inconsistencies are visible in the 
$\theta_\gamma$ ($\chi^2/\mbox{dof}=38/14$)
and $M_{2\pi}$ ($\chi^2/\mbox{dof}=17/11$) distributions. 
To estimate shifts in the efficiency correction due to
the dependence of the efficiency on these variables, we calculate 
\begin{equation}
 r_x =
 \sum_i\frac{P_\omega^{exp}(x_i)}{P_\omega^{MC}(x_i)}{P_\eta^{MC}(x_i)},
 \;x=\theta_\gamma, M_{2\pi},
\label{fcor}
\end{equation}
where $P_{\omega(\eta)}$ is the $\theta_\gamma$ or $M_{2\pi}$
distribution for $\omega\gamma$ ($\eta\gamma$) events normalized to
unit area, and $x_i$ is the center of the $i^{\rm th}$ bin.
We obtain the values 
$r_{\theta_\gamma}\! =\! 1.011\pm0.006$ and
$r_{M_{2\pi}}\! =\! 0.973\pm0.016$,
from which we calculate the efficiency correction 
$\delta_\eta\! =\! r_{\theta_\gamma} r_{M_{2\pi}}\delta_\omega\!
               =\! 0.918\pm0.032$,
and the detection efficiency
$\varepsilon\! =\!\delta_\eta\varepsilon_{MC}=(1.85 \pm 0.09)\%$.

\subsection{\boldmath $e^+e^-\to \eta^\prime\gamma$}

The simulated efficiency for $\eetoepg \!\!\to\! 2\pi 3\gamma$ 
events is $\varepsilon_{MC}\! =\! (2.91\pm0.13)\%$.
For this final state we can again use the efficiency
correction determined from the $e^+e^- \!\to\!\gisr\omega$ events, 
taking into account differences in the relevant kinematic variables.
Considering the same set of variables, we find similar results: 
corrections are needed only for $\theta_\gamma$ and $M_{2\pi}$ with
very similar values of 
$r_{\theta_\gamma}\! =\! 1.016\pm0.008$ and
$r_{M_{2\pi}}\! =\! 0.976\pm0.013$.
In addition, there are photon distributions that are different for $\pi^0$ and
$\eta$ decays.
We show distributions of the angle between the two decay photons
$\theta_{\gamma\gamma}$,
and the minimum and maximum photon energies $E_{\gamma, {\rm min}}$ and
$E_{\gamma, {\rm max}}$ in Fig.~\ref{effcor_etap}.
A disagreement between data and simulation for $\epem \!\!\to\! \gisr\omega$
events is seen in the $E_{\gamma, {\rm max}}$ spectrum
($\chi^2/\mbox{dof}=18/11$) and we calculate
$r_{E_{\gamma, {\rm max}}}\! =\! 1.035\pm0.016$.
This correction is not needed for \eetoeg events since their
$E_{\gamma, {\rm max}}$ distribution is very close to that for
$\epem \!\!\to\! \gisr\omega$ events.
We calculate an efficiency correction of
$\delta_{\eta^\prime}\!
 =\! r_{\theta_\gamma} r_{M_{2\pi}} r_{E_{\gamma, {\rm max}}} \delta_\omega\! 
 =\! 0.957\pm0.037$,
and a detection efficiency of
$\varepsilon\! =\!\delta_{\eta^\prime}\varepsilon_{MC}\!
               =\! (2.79 \pm 0.16)\%$.

The simulated efficiency for $\eetoepg \!\!\to\! 4\pi 3\gamma$ 
events is $(1.05 \pm 0.07)\%$.
We estimate an efficiency correction for the two additional pions
using the ISR process
$\epem \!\!\to\! \gisr\rho\eta \!\!\to\! \pi^+\pi^-\eta\gamma$.
We select events in both the $\eta \!\!\to\! \gamma\gamma$ and
$\eta \!\!\to\! \pip\pim\pi^0$ decay modes with criteria similar to those used
for the signal.
The $\pi^+\pi^-\eta$ invariant mass must be in the range
1.4--1.7~\gevcc where the $\rho\eta$ mass spectrum is at a maximum,
and the invariant mass of the $\rho$ candidate must be in the range
0.64--0.90~\gevcc.
From the numbers of selected data and simulated events in the two
$\eta$ decay modes we determine the double ratio
$\delta_{5\pi}\! =\! (N_{3\pi}/N_{2\gamma})_{data} /
                     (N_{3\pi}/N_{2\gamma})_{MC}\! =\! 0.98\pm$0.06,
and we calculate a fully corrected detection efficiency of
$\varepsilon\!
=\!\delta_{5\pi}\delta_{\eta^\prime}\varepsilon_{MC}\! =(0.99 \pm 0.10)\%$.
The ratio of the numbers of events selected in the two decay modes,
0.31$\pm$0.11, is consistent with the ratio of simulated detection
efficiencies 0.35$\pm$0.04.
The total detection efficiency for the two modes is $(3.78 \pm 0.19)\%$.

\section{Cross sections and form factors}

For each of the two signal processes, we calculate the cross section as
\begin{equation}
\sigma(e^+e^-\to P\gamma) = \frac{N_{P\gamma}}{\varepsilon L}R,
\label{crsec}
\end{equation}
where $N_{P\gamma}$ is the number of signal events from Sec.~\ref{bkgsub},
$\varepsilon$ is the detection efficiency from Sec.~\ref{deteff},
$L\! =$232~fb$^{-1}$ is the integrated luminosity,
and $R$ is a radiative correction factor.
We calculate $R$ as the ratio of the Born cross section for
$\epem \!\!\to\! P\gamma$ to the total cross section including
higher-order radiative corrections calculated with the structure
function method~\cite{strfun}.
The simulation requires the invariant mass of the $P\gamma$ system
$M_{P\gamma}\! >\,$8~\gevcc, for which we calculate $R\! =\,$0.956.
The detection efficiency used in Eq.(\ref{crsec}) is for simulated
events with this requirement.
The value of $R$ depends on the energy dependence of the cross section.
We use $\sigma \propto 1/q^{4}$ (see Eqs.~(\ref{eq1}) and~(\ref{eq3})),
and investigate the model dependence by recalculating
$R/\varepsilon$ under the $1/q^{3}$ and $1/q^{5}$ hypotheses.
The relative variation is less than $10^{-3}$, which we neglect. 
The theoretical uncertainty on $R$ obtained with the structure function
method does not exceed 1\%. 
We obtain
\begin{eqnarray}
\sigma(\epem \to \eta\gamma)
                           & = & 4.5^{+1.2}_{-1.1} \pm 0.3 \mbox{ fb},
\label{crsecres} \\
\sigma(\epem \to \eta^\prime\gamma) 
                           & = & 5.4       \pm 0.8 \pm 0.3 \mbox{ fb},
\label{crsecret}
\end{eqnarray}
where the first error is statistical and the second systematic.
The systematic error is the sum in quadrature of contributions
from detection efficiency, background subtraction,
fitting procedure, and radiative correction.

The value of $R$ we use does not take into account vacuum
polarization, and its contribution is included in the
results~(\ref{crsecres})--(\ref{crsecret}). 
For comparison with theoretical predictions, we calculate the 
so-called ``undressed'' cross section by applying a 7.5$\pm$0.2\%
correction for vacuum polarization at 10.58~\gevcc~\cite{vp}, obtaining
\begin{eqnarray}
\sigma(\epem \to \eta\gamma)_{\rm undressed}
                           & = & 4.2^{+1.2}_{-1.0} \pm 0.3 \mbox{ fb}, \\
\sigma(\epem \to \eta^\prime\gamma)_{\rm undressed}
                           & = & 5.0^{+0.8}_{-0.7} \pm 0.3 \mbox{ fb}.
\label{crsecres0}
\end{eqnarray}
Using Eq.(\ref{eq1}) we obtain the values of the $\eta\gamma$ and
$\eta^\prime\gamma$ transition form factors at $q^2=$112$\gev^2$
\begin{eqnarray}
q^2|F_\eta(q^2)|          & = & 0.229 \pm0.030 \pm0.008 \gev, \\
q^2|F_{\eta^\prime}(q^2)| & = & 0.251 \pm0.019 \pm0.008 \gev.
\end{eqnarray}

\begin{figure}
\includegraphics[width=0.8\linewidth]{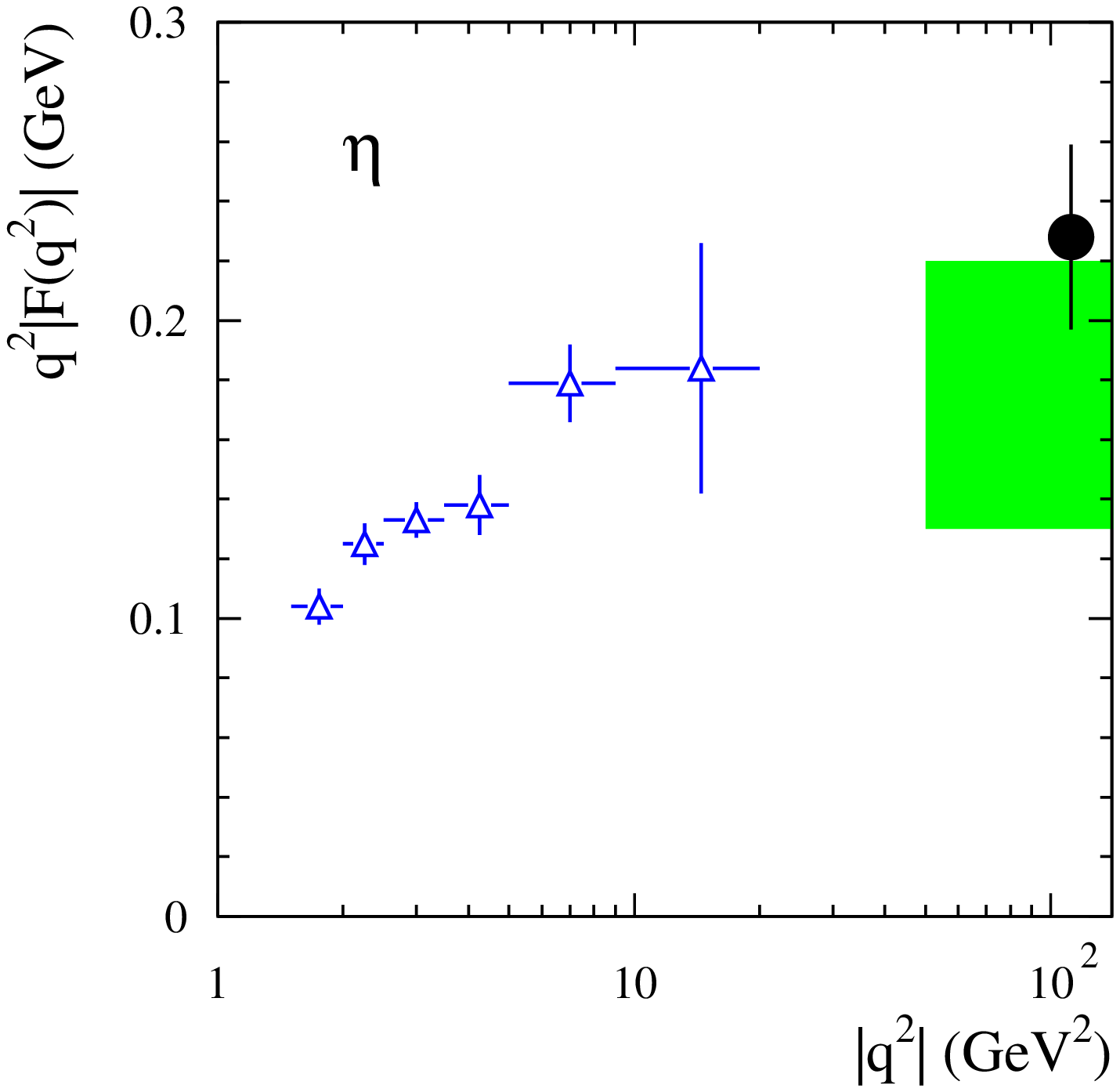}
\vspace{0.2cm}
\includegraphics[width=0.8\linewidth]{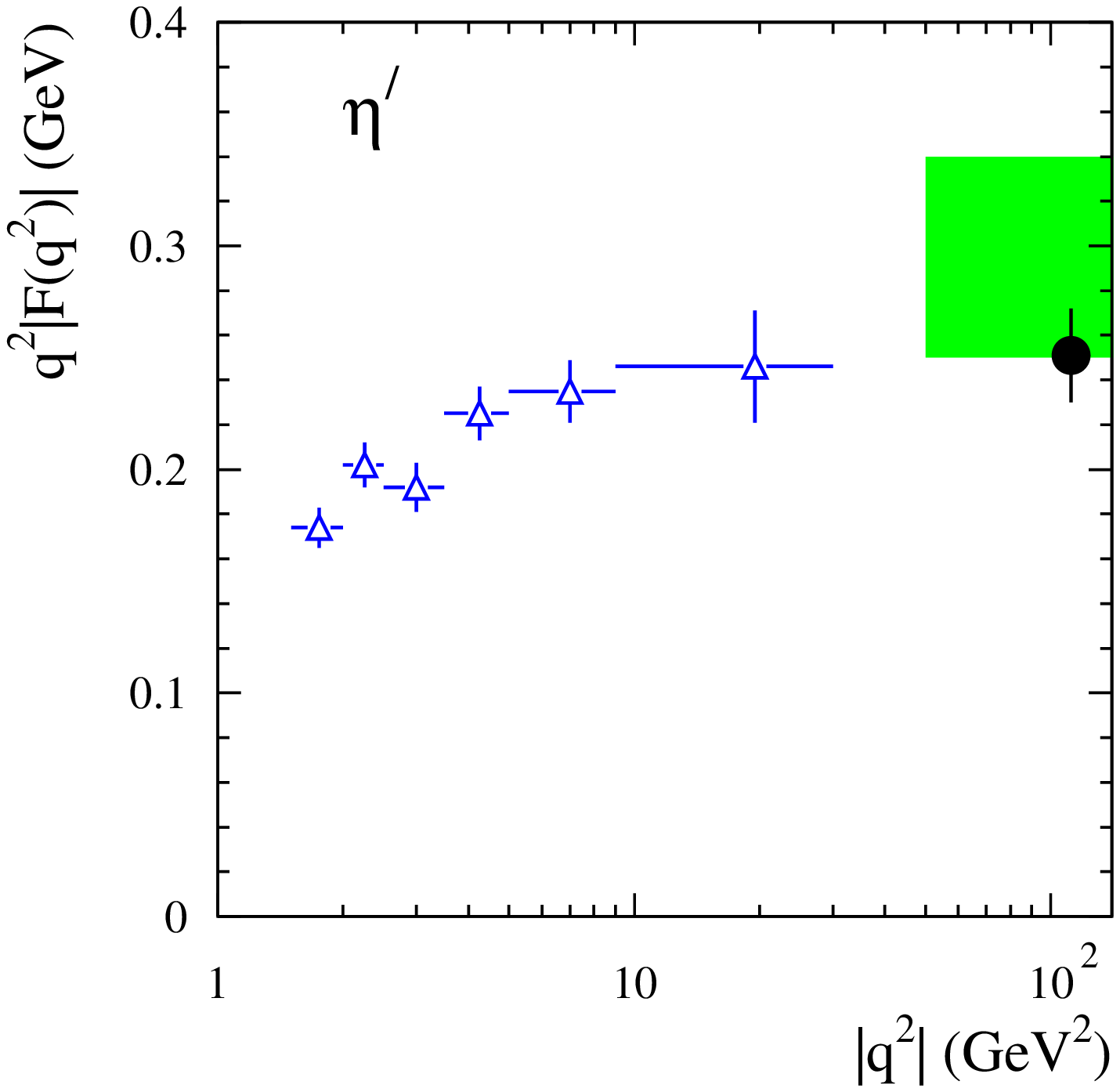}
\vspace{-0.2cm}
\caption{
  The magnitudes of the $\eta\gamma$ (top) and $\eta^\prime\gamma$
  (bottom) transition form factors measured in this work (filled circle) 
  and by CLEO~\cite{CLEO97} (triangles).
  The shaded boxes indicate the ranges of form-factor values
  calculated according to Eq.(\ref{eq3}) with the decay constants from 
  Refs.~\cite{Feldman,Kaiser,Benayoun,Goity,DeFazio,Escribano}.
\label{ff}}
\end{figure}

\section{Summary}

We have studied the \eetoeg and \eetoepg processes at an \epem c.m.\
energy  of 10.58 GeV.
We select $20^{+6}_{-5}$ $\eta\gamma$ and 
$50^{+8}_{-7}$ $\eta^\prime\gamma$ events, 
measure the cross sections and
extract the values of the transition form factors at $q^2\! =\! 112~\gev^2$.

Since the asymptotic values of the time-like and space-like
transition form factors are expected to be very close, we show our
results along with CLEO results for space-like momentum
transfers~\cite{CLEO97} in Fig.~\ref{ff} 
(we averaged
the CLEO results obtained in different $\eta$ ($\eta^\prime$) decay modes).
The CLEO data rise with increasing $q^2$, and are consistent with the
values given by our data points.  
A precise theoretical prediction of the value of the form factor 
at $q^2\! =\! 112~\gev^2$ is problematic due to uncertainties in the effective
decay constants, the quark distribution amplitudes, and possible gluon content
of the $\eta$ and $\eta^\prime$.  
Naively taking the decay constants from
Refs.~\cite{Feldman,Kaiser,Benayoun,Goity,DeFazio,Escribano} and calculating
form factor values according to Eq.(\ref{eq3}), we obtain
a range of values indicated by the shaded boxes in Fig.~\ref{ff}.  
Our data points are at the upper and lower ends of the range of predictions
for $\eta$ and $\eta^\prime$, respectively. 
The predicted ratio of the form factors ranges from 1.6 to 2.3, 
inconsistent with our value of $1.10 \pm 0.17$.  
This discrepancy
and the large range of the predictions indicates the need for more
theoretical input.

\section{ \boldmath Acknowledgments}
We thank V.L.~Chernyak, A.I.~Milstein and Z.K.~Silagadze for many 
fruitful discussions.
We are grateful for the 
extraordinary contributions of our \pep2\ colleagues in
achieving the excellent luminosity and machine conditions
that have made this work possible.
The success of this project also relies critically on the 
expertise and dedication of the computing organizations that 
support \babar.
The collaborating institutions wish to thank 
SLAC for its support and the kind hospitality extended to them. 
This work is supported by the
US Department of Energy
and National Science Foundation, the
Natural Sciences and Engineering Research Council (Canada),
Institute of High Energy Physics (China), the
Commissariat \`a l'Energie Atomique and
Institut National de Physique Nucl\'eaire et de Physique des Particules
(France), the
Bundesministerium f\"ur Bildung und Forschung and
Deutsche Forschungsgemeinschaft
(Germany), the
Istituto Nazionale di Fisica Nucleare (Italy),
the Foundation for Fundamental Research on Matter (The Netherlands),
the Research Council of Norway, the
Ministry of Science and Technology of the Russian Federation, and the
Particle Physics and Astronomy Research Council (United Kingdom). 
Individuals have received support from 
CONACyT (Mexico), the Marie-Curie Intra European Fellowship program (European Union),
the A. P. Sloan Foundation, 
the Research Corporation,
and the Alexander von Humboldt Foundation.

\end{document}